\documentclass{article}

\usepackage{PRIMEarxiv}

\usepackage[utf8]{inputenc} 
\usepackage[T1]{fontenc}    
\usepackage{hyperref}       
\usepackage{url}            
\usepackage{booktabs}       
\usepackage{amsmath}
\usepackage{amsthm}
\usepackage{amssymb}
\usepackage{amsfonts}
\usepackage{mathtools}
\usepackage{amsfonts}       
\usepackage{nicefrac}       
\usepackage{microtype}      
\usepackage{fancyhdr}       
\usepackage{graphicx}       
\usepackage{times}
\usepackage{subfig}
\usepackage{xcolor}
\usepackage{array}
\usepackage{cite}
\usepackage{multirow}
\usepackage{adjustbox}
\usepackage{fontawesome}
\usepackage[font=footnotesize,labelfont=bf]{caption}
\usepackage{tikz-network}
\usepackage{enumitem}

\DeclareMathOperator{\diag}{diag}
\DeclareMathOperator{\tr}{tr}

\newtheorem{theorem}{Theorem}
\newtheorem{lemma}{Lemma}
\newtheorem{proposition}{Proposition}
\newtheorem{corollary}{Corollary}

\newtheorem{assumption}{Assumption}

\newtheorem{remark}{Remark}

\title{
Distributed Adaptive Estimation of Unknown Nonlinear Systems without Input Sharing
\thanks{This work was supported by the Office of Naval Research under Grant N00014-21-1-2431 and the National Science Foundation under Grant 2208182.}
}

\author{
  Moh Kamalul Wafi and Milad Siami \\
  Department of Electrical and Computer Engineering \\
  Northeastern University \\
  Boston, MA, USA\\
  \texttt{\{wafi.m, m.siami\}@northeastern.edu} \\
}

\begin{document}
\maketitle

\begin{abstract}
This paper studies distributed adaptive state estimation for discrete-time nonlinear systems with unknown source dynamics over directed communication networks. Each sensing agent estimates the source state using only local measurements and information exchanged with neighboring agents, enabling a fully distributed implementation without requiring shared excitation or control inputs. A normalized adaptive estimation scheme is proposed to identify unknown linear and nonlinear dynamics while ensuring robust discrete-time adaptation. A Lyapunov-based analysis establishes input-to-state stability (ISS) of the estimation error dynamics, guaranteeing bounded adaptive parameters under bounded disturbances and asymptotic convergence of the estimation errors in the disturbance-free case under suitable conditions. To characterize the network-induced coupling, explicit norm-based and LMI-based Schur stability conditions are developed for the coupling operator, including a robust formulation accounting for bounded model uncertainty. Numerical simulations on star, cyclic, and path communication topologies demonstrate accurate distributed state estimation and validate the proposed stability conditions. Computational results further show that the proposed estimator scales efficiently with the network size.
\end{abstract}

\allowdisplaybreaks


\section{Introduction}
\label{sec:introduction}

Distributed state estimation over sensor networks is a fundamental problem in multi-agent systems, with applications in environmental monitoring, cooperative robotics, and intelligent transportation~\cite{R1}. In such settings, a network of sensing agents collaboratively estimates the state of a dynamic source using only local measurements and neighbor communication. Existing approaches can be broadly classified into consensus-based observers~\cite{R2,R3,R4}, which typically assume known source dynamics, and distributed Kalman filtering methods~\cite{R5,R6,R7}, which explicitly model stochastic disturbances but require accurate system models and noise statistics. In many practical applications, however, the source dynamics are only partially known or completely unknown~\cite{Wafi-Elham,R8,R9}. Although adaptive estimation techniques have been proposed for such settings~\cite{Wafi-Quadruple,R10,R11}, most existing methods primarily establish convergence under ideal conditions and provide limited robustness guarantees in the presence of bounded disturbances. Moreover, many distributed adaptive schemes rely on exchanging auxiliary excitation or input information among neighboring agents, which may be undesirable or impractical in bandwidth-limited or decentralized sensing networks.

Motivated by these limitations, this paper extends our previous work~\cite{Wafi-IFACWC}, which studied distributed adaptive estimation for partially known linear source dynamics in continuous and discrete time. Here, we address unknown nonlinear source dynamics in the discrete-time setting. The main new developments are the use of locally evaluated nonlinear regressors, a Kronecker-structured nonlinear error model, norm-based and LMI-based Schur stability certificates for the coupling operator, and a normalized adaptive law with Lyapunov guarantees for boundedness, convergence under suitable conditions, and ISS robustness. Similar idea to~\cite{R13}, the proposed estimator also avoids sharing excitation or control inputs among sensing nodes. Unlike existing distributed adaptive estimators, the proposed framework combines nonlinear adaptive estimation, a Kronecker-structured decomposition separating network and local dynamics, explicit Schur stability certificates for the coupling operator, and ISS guarantees under bounded disturbances within a unified discrete-time Lyapunov framework.

The main contributions of this paper are summarized as follows:
\begin{enumerate}[leftmargin=*]

\item A fully distributed adaptive estimation framework for discrete-time nonlinear systems with unknown source dynamics that does not require sensing nodes to exchange source or neighboring excitation inputs. The proposed formulation employs a Kronecker-product representation that separates the communication topology from the local observer dynamics.

\item An explicit spectral condition (Proposition~\ref{prop:NL:S_schur}) guaranteeing Schur stability of the network coupling operator through an inequality relating the communication topology, local observer contraction, and source dynamics.

\item A hierarchy of LMI-based Schur stability certificates consisting of a structured Lyapunov condition (Lemma~\ref{lem:NL:LMI_S}), a simplified scalar-weighted LMI (Corollary~\ref{cor:NL:practical_LMI}), and a robust extension accommodating norm-bounded model uncertainties (Corollary~\ref{cor:NL:robust_S}).

\item A Lyapunov-based stability analysis (Theorem~\ref{thm:AdpId:CL}) establishing boundedness of the adaptive estimator together with asymptotic convergence under suitable conditions. Furthermore, Corollaries~\ref{cor:AdpId:ParamBound}--\ref{cor:AdpId:ISS} establish bounded adaptive parameters and input-to-state stability (ISS) with respect to bounded predictor mismatch and external disturbances.

\item Extensive numerical studies on star, cyclic, and path communication topologies validating the theoretical results, together with scalability experiments demonstrating near-linear computational complexity with respect to the number of sensing nodes.

\end{enumerate}

\textbf{Notation.} 
$\mathbb{R}^n$ and $\mathbb{R}^{n\times p}$ denote real $n$-vectors and $n\times p$ matrices. $I_p$ is the $p\times p$ identity, and $\diag\{A_i\}\in\mathbb{R}^{pn\times pm}$ for $i=1,\dots,p$ forms a block-diagonal matrix with blocks $A_i\in\mathbb{R}^{n\times m}$. Vectors of ones and zeros in $\mathbb{R}^p$ are $\mathbf{1}_p$ and $\mathbf{0}_p$. The Kronecker product is $A\otimes B$, $\operatorname{tr}(A)$ is the trace, $\lambda(A)$ the eigenvalues, and $\operatorname{vec}(A)$ the vectorization of $A$.

\section{Communication Network}\label{sec:ComNetwork}

We consider a network consisting of \(m{+}1\) agents indexed by \(\mathcal{V}=\{0,1,\dots,m\}\), where agent \(0\) is the dynamic source to be estimated and agents \(1,\dots,m\) are sensing nodes. Each sensing node acts as a \emph{sensor--observer}: it collects local measurements, updates its own observer, and exchanges estimates with its in-neighbors to collaboratively estimate the source dynamics. The source itself is not part of the estimation network; it only generates the unknown state trajectory to be estimated.

Communication is modeled by a weighted directed graph \(\mathcal{G}=(\mathcal{V},\mathcal{E},\mathcal{W})\) with edge set \(\mathcal{E}\subseteq\mathcal{V}\times\mathcal{V}\) and nonnegative weights \(\mathcal{W}=\{w_{ij}\}\). An edge \((i,j)\in\mathcal{E}\) means agent \(i\) receives information from \(j\) with weight \(w_{ij}>0\); otherwise \(w_{ij}=0\). The in-neighborhood of agent \(i\) is \(\mathcal{N}_i \coloneqq \{\,j\in\mathcal{V}\mid(i,j)\in\mathcal{E}\,\}\). Here, the weights \(\{w_{ij}\}\) are algorithmic design coefficients (not physical signal strengths).

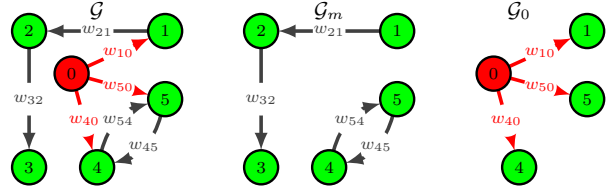
\begin{figure}[t!]
    \centering
    \scalebox{0.90}{{\begin{tikzpicture}
            \centering
            \Text[x=0,y=1.3,fontsize=\small]{$\mathcal{G}$};
            \Vertex[x=-.375,y=.375,label=$0$,color=red,opacity=0.1,size=.5]{L}
            \Vertex[x=1,y=1,label=$1$,color=green,opacity=0.1,size=.5]{1}
            \Vertex[x=-1,y=1,label=$2$,color=green,opacity=0.1,size=.5]{2}
            \Vertex[x=-1,y=-1,label=$3$,color=green,opacity=0.1,size=.5]{3}
            \Vertex[x=0,y=-1,label=$4$,color=green,opacity=0.1,size=.5]{4}
            \Vertex[x=1,y=0,label=$5$,color=green,opacity=0.1,size=.5]{5}
            \Edge[Direct,color=red,label=$w_{10}$](L)(1)
            \Edge[Direct,color=red,label=$w_{40}$](L)(4)
            \Edge[Direct,color=red,label=$w_{50}$](L)(5)
            \Edge[Direct,label=$w_{21}$](1)(2)
            \Edge[Direct,label=$w_{32}$](2)(3)
            \Edge[Direct,bend=30,label=$w_{45}$](5)(4)
            \Edge[Direct,bend=30,label=$w_{54}$](4)(5)
        \end{tikzpicture}}}
    \qquad
    \scalebox{0.90}{{\begin{tikzpicture}
            \centering
            \Text[x=0,y=1.3,fontsize=\small]{$\mathcal{G}_m$};
            \Vertex[x=1,y=1,label=$1$,color=green,opacity=0.1,size=.5]{1}
            \Vertex[x=-1,y=1,label=$2$,color=green,opacity=0.1,size=.5]{2}
            \Vertex[x=-1,y=-1,label=$3$,color=green,opacity=0.1,size=.5]{3}
            \Vertex[x=0,y=-1,label=$4$,color=green,opacity=0.1,size=.5]{4}
            \Vertex[x=1,y=0,label=$5$,color=green,opacity=0.1,size=.5]{5}
            \Edge[Direct,label=$w_{21}$](1)(2)
            \Edge[Direct,label=$w_{32}$](2)(3)
            \Edge[Direct,bend=30,label=$w_{45}$](5)(4)
            \Edge[Direct,bend=30,label=$w_{54}$](4)(5)
        \end{tikzpicture}}}
    \qquad
    \scalebox{0.90}{{\begin{tikzpicture}
            \centering
            \Text[x=0,y=1.3,fontsize=\small]{$\mathcal{G}_0$};
            \Vertex[x=-.375,y=.375,label=$0$,color=red,opacity=0.1,size=.5]{L}
            \Vertex[x=1,y=1,label=$1$,color=green,opacity=0.1,size=.5]{1}
            \Vertex[x=0,y=-1,label=$4$,color=green,opacity=0.1,size=.5]{4}
            \Vertex[x=1,y=0,label=$5$,color=green,opacity=0.1,size=.5]{5}
            \Edge[Direct,color=red,label=$w_{10}$](L)(1)
            \Edge[Direct,color=red,label=$w_{40}$](L)(4)
            \Edge[Direct,color=red,label=$w_{50}$](L)(5)
        \end{tikzpicture}}}
    \caption{Example of a graph $\mathcal{G}$ with $m=5$, sensing-to-sensing subgraph $\mathcal{G}_m$, and  source-to-sensing subgraph $\mathcal{G}_0$, showing the decoupling and assignment of $w_{ij}$.}
    \label{Fig:network}
\end{figure}

For analysis, we separate \(\mathcal{G}\) into two induced subgraphs:
(i) the sensing-only subgraph \(\mathcal{G}_m=(\mathcal{V}_m,\mathcal{E}_m,\mathcal{W}_m)\) with \(\mathcal{V}_m=\{1,\dots,m\}\);
(ii) the source--sensing subgraph \(\mathcal{G}_0=(\mathcal{V}_0,\mathcal{E}_0,\mathcal{W}_0)\) with \(\mathcal{V}_0=\{0\}\cup\{\,i\in\mathcal{V}_m\mid(i,0)\in\mathcal{E}\,\}\).
Thus, \(\mathcal{E}_m\) contains sensing--sensing links and \(\mathcal{E}_0\) contains links from the source to sensing nodes (see Fig.~\ref{Fig:network}). This decomposition separates inter-sensor communication from source information, enabling a compact Kronecker representation for the stability analysis.

For \(\mathcal{G}_m\), define the in-degree $d_i \coloneqq \sum_{j:(i,j)\in\mathcal{E}_m} w_{ij}$, $i\in\mathcal{V}_m$ and \(\mathbb{D}_m=\diag\{d_1,\dots,d_m\}\).
The adjacency and Laplacian of the sensing subgraph are $[\mathbb{A}_m]_{ij} = w_{ij}$ if $(i,j)\in\mathcal{E}_m$ (with 0 otherwise) and $\mathbb{L}_m \coloneqq \mathbb{D}_m - \mathbb{A}_m$.
Regarding \(\mathcal{G}_0\), it is encoded by a diagonal matrix $\mathbb{A}_0 \coloneqq \diag\{w_{10},\dots,w_{m0}\}$ where $[\mathbb{A}_0]_{ii} = w_{i0}$ if $(i,0)\in\mathcal{E}_0$ and 0 otherwise.

Let \(\mathbb{W}\coloneqq \diag\{w_1,\dots,w_m\}\) with \(w_i\coloneqq d_i + w_{i0}\). The network coupling matrix is defined as
\begin{equation*}
    \mathbb{L} \coloneqq \mathbb{L}_m + \mathbb{A}_0 = \mathbb{W} - \mathbb{A}_m.
\end{equation*}
We propose the \emph{balanced} case \(\mathbb{W}=I_m\), which implies
\begin{equation}\label{eq:ComNet:balance}
    (\mathbb{L}-\mathbb{A}_0)\mathbf{1}_m=\mathbf{0}_m
    \quad\Leftrightarrow\quad
    (\mathbb{A}_m+\mathbb{A}_0)\mathbf{1}_m=\mathbf{1}_m,
\end{equation}
i.e., each sensing node’s total incoming weight from neighbors and the source sums to one.\footnote{If \(\mathbb{W}\neq I_m\), normalization via \(\tilde w_{ij}\coloneqq w_{ij}/w_i\) for \((i,j)\in\mathcal{E}_m\) and \(\tilde w_{i0}\coloneqq w_{i0}/w_i\) yields \(\tilde{\mathbb{W}}=I_m\) and preserves \eqref{eq:ComNet:balance} (with \(\tilde{\mathbb{L}}=\tilde{\mathbb{L}}_m+\tilde{\mathbb{A}}_0\)); this is not required for implementation but simplifies stability analysis.}

\begin{remark}[Source reachability]\label{rem:reachability}
At least one sensing node directly receives information from the source (\(w_{i0}>0\) for some \(i\)), and every sensing node is reachable from the source via a directed path. Under this assumption, the network coupling matrix \(\mathbb{L}=\mathbb{L}_m+\mathbb{A}_0\) is \emph{positive stable}, i.e., \(\Re\{\lambda(\mathbb{L})\}>0\).
\end{remark}

\section{Problem Formulation}\label{sec:ProblemFormulation}

We consider a distributed sensor network composed of an \emph{unknown} nonlinear source, indexed by $0$, and $m$ sensing nodes, indexed by $i\in\{1,\dots,m\}$. Each sensing node collects local information and executes an adaptive observer. 

The unknown source evolves as
\begin{equation}
    x_{0,k+1} = A_0 x_{0,k} + f_{0}(x_{0,k})  + B_0 u_{0,k} + \delta_{0,k}, \label{eq:Problem:source} 
\end{equation}
where $x_{0,k}\in\mathbb{R}^n$ represents the state and $u_{0,k}\in\mathbb{R}^p$ is the \emph{unknown} source input. The matrix $A_0$ is assumed to be \emph{Schur stable} and is \emph{unknown} whereas $B_0$ is \emph{known}. Here, $A_0$ can be decomposed as $A_0 = A_\ast + \Delta A$, where $A_\ast$ is known and the uncertainty $\Delta A$ satisfies $\|\Delta A\|_2\le a$ for some known constant $a>0$. 
The nonlinear function $f_0(\cdot)$ is expressed as $f_{0}(x_{0,k})\coloneqq C_0g(x_{0,k})$ where $C_0\in\mathbb R^{n\times r}$ is an \emph{unknown} matrix and $g:\mathbb R^n\to\mathbb R^r$ is a known smooth nonlinear mapping. 
The disturbance $\delta_{0,k}\in\mathbb R^n$ is bounded with $\|\delta_{0,k}\|\le \delta^\ast$ for some $\delta^\ast>0$.

\begin{assumption}\label{assmpt:Lipschitz}
    The mapping $g:\mathbb{R}^n\to\mathbb{R}^r$ is Lipschitz continuous; that is, there exists a constant $L_g>0$ such that
    \begin{equation}\label{eq:g_Lipschitz}
        \|g(x)-g(y)\|
        \le L_g\|x-y\|,
        \qquad \forall x,y\in\mathbb{R}^n.
    \end{equation}
\end{assumption}

The matrices $A_0$ and $C_0$, together with the source input $u_{0,k}$,
are assumed unknown to all sensing nodes. The objective is to estimate
the source dynamics using only locally available information and
neighbor communications.

Sensing node $i\in\{1,\dots,m\}$ implements
\begin{subequations}
\begin{align}
    x_{i,k+1} &= S_i x_{i,k} + (\hat{A}_{i,k} - S_i) z_{i,k} 
    + \hat{C}_{i,k}g(z_{i,k}) + B_{i} u_{i,k}, \label{eq:Problem:sensor} \\
    z_{i,k} &= \sum\nolimits_{j\in\mathcal{N}_{i}} w_{ij}x_{j,k}, \qquad \mathcal{N}_{i}\subseteq\mathcal{V} \label{eq:Problem:z}
\end{align}
where $x_{i,k}\in\mathbb{R}^n$ is the state estimate and $z_{i,k}$ is the aggregated estimate obtained from its in-neighbors
\footnote{The nonlinear term $g(\cdot)$ in \eqref{eq:Problem:sensor} is evaluated using the locally available aggregated estimate $z_{i,k}$ rather than the true source state.}.
The matrices $\hat{A}_{i,k}\in\mathbb R^{n\times n}$ and $\hat{C}_{i,k}\in\mathbb R^{n\times r}$ are the estimates of the \emph{unknown} $A_{0}$ and $C_0$, respectively.  
All sensing nodes use a common known input matrix $B_i = B_0$, while each $S_i\in\mathbb R^{n\times n}$ is a user-designed \emph{Schur stable} matrix. Each sensing node updates its local input according to
\begin{equation}\label{eq:Problem:u_update}
    u_{i,k+1} \!=\! u_{i,k} - \underbrace{\frac{\gamma_{u_i}}{N_{i,k}} B_i^\top \bigl(P_i e_{i,k} + \sum\nolimits_{j\in\mathcal{N}_i} [\mathbb{L}]_{ji} P_j e_{j,k}\bigr)}_{h_i(\cdot)},
\end{equation}
\end{subequations}
where $\gamma_{u_i}>0$, $N_{i,k}$ is a normalization factor, $P_i\succ0$ is a weighting matrix and $e_{i,k}\coloneqq x_{i,k}-z_{i,k}$ denotes the local estimation error.  
Notice that the update law \eqref{eq:Problem:u_update} is implemented locally, \emph{without} requiring the source input $u_{0,k}$ or the neighboring inputs $u_{j,k}$.

\begin{remark}[Locally available information]\label{rem:local}
    Each sensing node has access only to its own estimate $x_{i,k}$, the aggregated estimate $z_{i,k}$, the known nonlinear mapping $g(\cdot)$, and the common input matrix $B_0$. The source input $u_{0,k}$ and neighboring inputs $u_{j,k}$ are never communicated. Consequently, every observer update is computed solely from locally available information.
\end{remark}

\textbf{Main objective}: Design distributed adaptive laws for
$\hat{A}_{i,k}$, $\hat{C}_{i,k}$, and the local input $u_{i,k}$ such that,
under the \emph{balanced} condition \eqref{eq:ComNet:balance} and Remark~\ref{rem:reachability},
both the estimation error
$e_{i,k}\coloneqq x_{i,k}-z_{i,k}$
and the source-tracking error
$\epsilon_{i,k}\coloneqq x_{i,k}-x_{0,k}$
are ISS with respect to the adaptive predictor mismatch and the bounded disturbance in
\eqref{eq:Problem:source}. Specifically, the objective is to guarantee
\begin{align}\label{eq:ProbFor:dis:ISS}
    \begin{aligned}
    \limsup\nolimits_{k\to\infty}\|e_{i,k}\|
    &\le
    \vartheta_{e,\eta}\eta^\ast
    +
    \vartheta_{e,\delta}\delta^\ast,\\
    \limsup\nolimits_{k\to\infty}\|\epsilon_{i,k}\|
    &\le
    \vartheta_{\epsilon,\eta}\eta^\ast
    +
    \vartheta_{\epsilon,\delta}\delta^\ast,
    \end{aligned}
\end{align}
for all $i\in\{1,\dots,m\}$ and some constants
$\vartheta_{e,\eta},\vartheta_{e,\delta},
\vartheta_{\epsilon,\eta},\vartheta_{\epsilon,\delta}>0$.
In the disturbance-free case
$(\eta^\ast=\delta^\ast=0)$,
both errors converge to zero:
$\lim_{k\to\infty}\|e_{i,k}\|=0$
and
$\lim_{k\to\infty}\|\epsilon_{i,k}\|=0$
for all $i$.

We end this section by defining shorthand notations utilized in the following sections, to avoid repetition. For any source signal $\beta_{0,k}\in\mathbb{R}^{q}$ and any sensing node signal $\beta_{i,k}\in\mathbb{R}^{q}$,
\begin{subequations}
\begin{align}
    \Bar{\beta}_{0,k} &= \mathbf{1}_m \otimes \beta_{0,k} \in \mathbb{R}^{mq}, \label{eq:ProbFor:beta0:k}\\
    \Bar{\beta}_{k} &= [\beta_{1,k}^\top,\dots,\beta_{m,k}^\top]^\top \in \mathbb{R}^{mq}. 
    \label{eq:ProbFor:betai:k}
\end{align}
\end{subequations}
For any matrices $E_0, E_i, \hat{E}_{i,k}\in\mathbb{R}^{q_1\times q_2}$,
\begin{subequations}
\begin{align}
    \mathbf{E}_0 &= I_m \otimes E_0 \in\mathbb{R}^{mq_1\times mq_2}, \label{eq:ProbFor:mat0}\\
    \mathbf{E} &= \diag\{E_1,\dots,E_m\}\in\mathbb{R}^{mq_1\times mq_2}, \label{eq:ProbFor:mati}\\
    \hat{\mathbf{E}}_k &= \diag\{\hat{E}_{1,k},\dots,\hat{E}_{m,k}\}\in\mathbb{R}^{mq_1\times mq_2}. \label{eq:ProbFor:mati:k}
\end{align}
\end{subequations}
With these definitions, we are ready to study the adaptive laws and their stability in networked discrete-time systems.

\section{Distributed Adaptive Estimation}\label{sec:AdaptiveId}

This section develops the distributed adaptive estimation framework. We first derive the network-level error dynamics, then establish sufficient conditions for the Schur stability of the coupling operator. Finally, distributed adaptive laws are proposed and their stability is analyzed using a Lyapunov approach.

\subsection{Error Dynamics in Networked Systems}\label{subsec:AdpId:Error}
In network-level form, the source evolves as
\begin{equation}
    \Bar{x}_{0,k+1} = \mathbf{A}_0 \Bar{x}_{0,k} + \mathbf{C}_0\mathbf{g}(\Bar{x}_{0,k}) + \mathbf{B}_0 \Bar{u}_{0,k} + \Bar{\delta}_{0,k}, \label{eq:AdpId:source}
\end{equation}
and the sensing node dynamics implement
\begin{subequations}
\begin{align}
    \Bar{x}_{k+1} &= \mathbf{S}\Bar{x}_{k} + (\hat{\mathbf{A}}_{k} - \mathbf{S})\Bar{z}_{k} + \hat{\mathbf{C}}_{k}\mathbf{g}(\Bar{z}_k) + \mathbf{B}\Bar{u}_{k}, \label{eq:AdpId:sensor} \\
    \Bar{z}_{k} &= (\mathbb{A}_m\otimes I_n)\Bar{x}_{k} + (\mathbb{A}_0\otimes I_n)\Bar{x}_{0,k}, \label{eq:AdpId:z} \\
    \Bar{u}_{k+1} &= \Bar{u}_k - \mathbf{h}(\cdot),
    \label{eq:AdpId:u}
\end{align}
\end{subequations}
The quantities $\Bar{x}_{0,k}$, $\Bar{u}_{0,k}$, and $\Bar{\delta}_{0,k}$ are defined via \eqref{eq:ProbFor:beta0:k}, while $\Bar{x}_{k}$, $\Bar{z}_{k}$, and $\Bar{u}_{k}$ use the stacking in \eqref{eq:ProbFor:betai:k}. The matrices $\mathbf{A}_0$, $\mathbf{B}_0$, and $\mathbf{C}_0$ follow \eqref{eq:ProbFor:mat0} while the matrix $\mathbf{S}$ follows \eqref{eq:ProbFor:mati}. The estimate matrices $\hat{\mathbf{A}}_{k}$ and $\hat{\mathbf{C}}_{k}$ are denoted as in \eqref{eq:ProbFor:mati:k}. The input matrix $\mathbf{B}$ is defined as $\mathbf{B} \equiv \mathbf{B}_0$ while the nonlinear terms are represented as
$\mathbf g(\Bar x_{0,k})=\mathbf 1_m\otimes g(x_{0,k})$,
$\mathbf{g}(\Bar{z}_{k})=[g(z_{1,k})^\top, \dots, g(z_{m,k})^\top]^{\top}$ and 
$\mathbf{h}(\cdot) = [h_1(\cdot)^\top, \dots, h_m(\cdot)^\top]^\top$. The network weighting matrices $\mathbb{A}_m$ and $\mathbb{A}_0$ in \ref{eq:AdpId:z} are defined in Section~\ref{sec:ComNetwork}. 

The derivation of the error dynamics is inspired by~\cite{Wafi-MRAC,Wafi-LCSS24}. For notational convenience, define the ideal parameter matrices
\begin{equation}\label{eq:AdpId:ideal_param}
    \hat{\mathbf{A}}^\ast\coloneqq\mathbf{A}_0,
    \qquad
    \hat{\mathbf{C}}^\ast\coloneqq\mathbf{C}_0.
\end{equation}
Define the stacked source-tracking error as
\begin{equation}\label{eq:AdpId:tracking_error}
    \Bar{\epsilon}_k
    \coloneqq
    \Bar{x}_k-\Bar{x}_{0,k}.
\end{equation}
Under the balanced condition \eqref{eq:ComNet:balance}, the stacked estimation error is
\begin{subequations}
\begin{align}
    \Bar e_k \coloneqq \Bar{x}_{k} - \Bar{z}_{k}
    &=(\mathbb L\otimes I_n)\Bar x_k-(\mathbb A_0\otimes I_n)\Bar x_{0,k} 
    \label{eq:AdpId:error} \\
    &=(\mathbb L\otimes I_n)(\Bar x_k-\Bar x_{0,k}) \nonumber \\
    &=(\mathbb L\otimes I_n)\Bar\epsilon_k, \label{eq:AdpId:error_tracking_relation}
\end{align}
\end{subequations}
where we used
$(\mathbb L\otimes I_n)\Bar x_{0,k}=(\mathbb A_0\otimes I_n)\Bar x_{0,k}.$
Since $\mathbb L$ is nonsingular under Remark~\ref{rem:reachability},
\begin{equation}\label{eq:AdpId:tracking_from_error}
    \Bar\epsilon_k
    =
    (\mathbb L^{-1}\otimes I_n)\Bar e_k.
\end{equation}
Moreover, using \eqref{eq:AdpId:z} and
$(\mathbb A_m+\mathbb A_0)\mathbf 1_m=\mathbf 1_m$,
\begin{equation}\label{eq:AdpId:z_tracking_relation}
\begin{aligned}
    \Bar z_k-\Bar x_{0,k}
    &=(\mathbb{A}_m\otimes I_n)\Bar{x}_{k} + (\mathbb{A}_0\otimes I_n)\Bar{x}_{0,k} - \Bar{x}_{0,k} \\
    &=(\mathbb A_m\otimes I_n)(\Bar x_k-\Bar x_{0,k}) \\
    &=(\mathbb A_m\mathbb L^{-1}\otimes I_n)\Bar e_k.
\end{aligned}
\end{equation}

Next, introduce the parameter errors, input mismatch, and nonlinear mapping mismatch variables
\begin{equation}\label{eq:AdpId:par_error}
    \begin{aligned}
    \Phi_k&\coloneqq\hat{\mathbf A}_k-\mathbf A_0,
    &
    \Psi_k&\coloneqq\hat{\mathbf C}_k-\mathbf C_0,
    \\
    \Bar r_k&\coloneqq\Bar u_k-\Bar u_{0,k},
    &
    \Delta\mathbf g_k&\coloneqq\mathbf g(\Bar z_k)-\mathbf g(\Bar x_{0,k}).
    \end{aligned}
\end{equation}
Also let $\Omega\coloneqq\mathbb L\otimes I_n.$
From \eqref{eq:AdpId:error}, the estimation error at time $k+1$ satisfies
\begin{align}\label{eq:AdpId:error_step1}
    \begin{aligned}
    \Bar e_{k+1}
    &=
    \Omega\Bar x_{k+1}
    -
    (\mathbb A_0\otimes I_n)\Bar x_{0,k+1} \\
    &=
    \Omega
    \bigl[
        \mathbf S\Bar x_k
        +
        (\hat{\mathbf A}_k-\mathbf S)\Bar z_k
        +
        \hat{\mathbf C}_k\mathbf g(\Bar z_k)
        +
        \mathbf B_0\Bar u_k
    \bigr] \\
    &\quad
    -
    (\mathbb A_0\otimes I_n)
    \bigl[
        \mathbf A_0\Bar x_{0,k}
        +
        \mathbf C_0\mathbf g(\Bar x_{0,k})
        +
        \mathbf B_0\Bar u_{0,k}
        +
        \Bar\delta_{0,k}
    \bigr].
    \end{aligned}
\end{align}
For any replicated source quantity $\Bar \beta_{0,k}=\mathbf 1_m\otimes \beta_{0,k}$,
the balanced condition gives
$(\mathbb A_0\otimes I_q)\Bar \beta_{0,k}=(\mathbb L\otimes I_q)\Bar \beta_{0,k}.$
Therefore, \eqref{eq:AdpId:error_step1} becomes
\begin{align}\label{eq:AdpId:error_step2}
    \begin{aligned}
    \Bar e_{k+1}
    &=
    \Omega
    \bigl[
        \mathbf S(\Bar x_k-\Bar z_k)
        +
        \hat{\mathbf A}_k\Bar z_k
        -
        \mathbf A_0\Bar x_{0,k}
    \bigr]
    +
    \Omega
    \bigl[
        \hat{\mathbf C}_k\mathbf g(\Bar z_k)
        -
        \mathbf C_0\mathbf g(\Bar x_{0,k})
    \bigr] \\
    &\quad
    +
    \Omega\mathbf B_0
    (\Bar u_k-\Bar u_{0,k})
    -
    \Omega\Bar\delta_{0,k}.
    \end{aligned}
\end{align}
Using 
$\Delta\mathbf g_k=\mathbf g(\Bar z_k)-\mathbf g(\Bar x_{0,k})$,
$\hat{\mathbf A}_k=\mathbf A_0+\Phi_k$, and 
$\hat{\mathbf C}_k=\mathbf C_0+\Psi_k$ give
\begin{align}\label{eq:AdpId:AC_decomposition}
    \begin{aligned}
    \hat{\mathbf A}_k\Bar z_k
    -
    \mathbf A_0\Bar x_{0,k}
    &=
    \mathbf A_0
    (\Bar z_k-\Bar x_{0,k})
    +
    \Phi_k\Bar z_k \\
    \hat{\mathbf C}_k\mathbf g(\Bar z_k)
    -
    \mathbf C_0\mathbf g(\Bar x_{0,k})
    &=
    \Psi_k\mathbf g(\Bar z_k)
    +
    \mathbf C_0\Delta\mathbf g_k.
    \end{aligned}
\end{align}
Using $\Bar e_k=\Bar x_k-\Bar z_k$ and
substituting \eqref{eq:AdpId:AC_decomposition} into
\eqref{eq:AdpId:error_step2} yields
\begin{align}\label{eq:AdpId:error_step3}
    \begin{aligned}
    \Bar e_{k+1}
    &=
    \Omega\mathbf S\Bar e_k
    +
    \Omega\mathbf A_0
    (\Bar z_k-\Bar x_{0,k})
    +
    \Omega
    \bigl[
        \Phi_k\Bar z_k
        +
        \Psi_k\mathbf g(\Bar z_k)
        +
        \mathbf B_0\Bar r_k
    \bigr] \\
    &\quad
    +
    \Omega\mathbf C_0\Delta\mathbf g_k
    -
    \Omega\Bar\delta_{0,k}.
    \end{aligned}
\end{align}
Using \eqref{eq:AdpId:z_tracking_relation} and the mixed-product property of the Kronecker product,
$(A\otimes B)(C\otimes D)=(AC)\otimes(BD),$
we obtain
\begin{align*}
    \Omega\mathbf A_0
    (\Bar z_k-\Bar x_{0,k})
    &=
    (\mathbb L\otimes I_n)
    (I_m\otimes A_0)
    (\mathbb A_m\mathbb L^{-1}\otimes I_n)
    \Bar e_k \nonumber\\
    &=
    (\mathbb L\otimes A_0)
    (\mathbb A_m\mathbb L^{-1}\otimes I_n)
    \Bar e_k \nonumber\\
    &=
    (\mathbb L\mathbb A_m\mathbb L^{-1}\otimes A_0)
    \Bar e_k.
\end{align*}
Under the balanced condition \eqref{eq:ComNet:balance},
$\mathbb A_m=I_m-\mathbb L,$ which implies
\[
    \mathbb L\mathbb A_m
    =\mathbb L(I_m-\mathbb L)
    =(I_m-\mathbb L)\mathbb L
    =\mathbb A_m\mathbb L.
\]
Hence, $\mathbb L\mathbb A_m\mathbb L^{-1}=\mathbb A_m,$
and therefore
\begin{align}\label{eq:AdpId:A_error_relation}
    \begin{aligned}
    \Omega\mathbf A_0
    (\Bar z_k-\Bar x_{0,k})
    &=(\mathbb A_m\otimes A_0)\Bar e_k \\
    &=(\mathbb A_m\otimes I_n)(I_m\otimes A_0)\Bar e_k \\
    &=(\mathbb A_m\otimes I_n)\mathbf A_0\Bar e_k.
    \end{aligned}
\end{align}
Hence, the complete estimation-error dynamics are
\begin{equation}\label{eq:AdpId:errordyn}
    \Bar e_{k+1}
    =
    \mathcal S\Bar e_k
    +
    \Omega\Bar\eta_k
    +
    \Omega\mathbf C_0\Delta\mathbf g_k
    -
    \Omega\Bar\delta_{0,k},
\end{equation}
where
\begin{subequations}
\begin{align}
    \mathcal S
    &\coloneqq\Omega\mathbf S
    +(\mathbb A_m\otimes I_n)\mathbf A_0 \label{eq:AdpId:S}\\
    \Bar\eta_k 
    &\coloneqq \Phi_k\Bar z_k
    +\Psi_k\mathbf g(\Bar z_k)
    +\mathbf B_0\Bar r_k. \label{eq:AdpId:eta}
\end{align}
\end{subequations}
Equivalently, $\Bar\eta_k=\Xi_k\Bar\xi_k$
with $\Xi_k=[\Phi_k,\Psi_k,\mathbf B_0]$ and
$\Bar\xi_k=[\Bar z_k^\top,\mathbf g(\Bar z_k)^\top,\Bar r_k^\top]^\top.$
Notice that by Assumption~\ref{assmpt:Lipschitz}, the nonlinear mismatch satisfies
\begin{align}
    \|\Delta\mathbf g_k\|^2
    &=
    \sum_{i=1}^m
    \|g(z_{i,k})-g(x_{0,k})\|^2
    \nonumber\\
    &\le
    L_g^2
    \sum_{i=1}^m
    \|z_{i,k}-x_{0,k}\|^2
    \nonumber\\
    &=
    L_g^2
    \|\Bar z_k-\Bar x_{0,k}\|^2
    \nonumber\\
    &\le
    L_g^2
    \|\mathbb A_m\mathbb L^{-1}\otimes I_n\|^2
    \|\Bar e_k\|^2,
    \label{eq:AdpId:g_mismatch_bound}
\end{align}
where the last inequality follows from
$\Bar e_k=(\mathbb L\mathbb A_m^{-1}\otimes I_n)(\Bar z_k-\Bar x_{0,k})$.
Before analyzing the stability of \eqref{eq:AdpId:errordyn}, we next discuss the operator $\mathcal{S}$ in \eqref{eq:AdpId:S}.

\subsection{Stability of the Coupling Operator $\mathcal{S}$}\label{subsec:AdpId:S}

To guarantee exponential convergence in the discrete-time setting, we now establish sufficient norm-based conditions under which the coupling operator $\mathcal{S}$ is Schur stable.

\begin{proposition}\label{prop:NL:S_schur}
Let $\mathcal{S}$ be defined by \eqref{eq:AdpId:S}. If
\begin{equation}\label{eq:NL:S_schur:cond}
    \|\mathbb{L}\|_2
    \max_{i}\|S_i\|_2
    +
    \|\mathbb{A}_m\|_2
    \|A_0\|_2
    <1,
\end{equation}
then $\mathcal{S}$ is Schur stable; that is,
$\rho(\mathcal{S})<1$.
\end{proposition}

Condition~\eqref{eq:NL:S_schur:cond} characterizes the interplay between the communication network and the local dynamics. The network gains $\|\mathbb{L}\|_2$ and $\|\mathbb{A}_m\|_2$ scale the contributions of the observer dynamics and the source dynamics, respectively. Consequently, stronger local contraction (smaller $\max_i\|S_i\|_2$) or weaker network gains can compensate for a larger induced norm $\|A_0\|_2$ of the source dynamics, ensuring that $\mathcal{S}$ remains Schur stable.

While Proposition~\ref{prop:NL:S_schur} provides an easily verifiable sufficient condition, it may be conservative. Lemma~\ref{lem:NL:LMI_S} provides a structured LMI certificate based on a Kronecker-product Lyapunov function that retains the full matrix structure of $\mathcal{S}$. Corollary~\ref{cor:NL:practical_LMI} is obtained by applying the weighted Young inequality to bound the cross terms in the structured Lyapunov inequality of Lemma~\ref{lem:NL:LMI_S}, yielding a simplified but more conservative condition.

\begin{lemma}[Structured Lyapunov LMI]\label{lem:NL:LMI_S}
Let $\mathcal{S}$ be defined by \eqref{eq:AdpId:S}, and fix a matrix
$P_n=P_n^\top\succ0$. If there exists a matrix
$P_m=P_m^\top\succ0$ such that, with
$P_{mn}\coloneqq P_m\otimes P_n$,
\begin{subequations}
\begin{equation}\label{eq:NL:structured_LMI}
    \begin{bmatrix}
        P_{mn} & \mathcal{S}^\top P_{mn} \\
        P_{mn}\mathcal{S} & P_{mn}
    \end{bmatrix}
    \succ0,
\end{equation}
then $\mathcal{S}$ is Schur stable. Equivalently,
\begin{equation}\label{eq:NL:structured_Lyap}
    \mathcal{S}^\top P_{mn}\mathcal{S}-P_{mn}\prec0.
\end{equation}
\end{subequations}
Moreover, the quadratic function $V_k=\Bar e_k^\top P_{mn}\Bar e_k$
satisfies $V_{k+1}-V_k<0$ for every $\Bar e_k\neq0$ along
$\Bar e_{k+1}=\mathcal{S}\Bar e_k$.
\end{lemma}

\begin{corollary}[Simplified structured LMI]\label{cor:NL:practical_LMI}
Let
\begin{subequations}
\begin{equation}\label{eq:NL:XY}
    X\coloneqq(\mathbb{L}\otimes I_n)\mathbf{S},
    \qquad
    Y\coloneqq(\mathbb{A}_m\otimes I_n)\mathbf{A}_0,
\end{equation}
so that $\mathcal{S}=X+Y$. Fix $P_n\succ0$ and a scalar
$\varepsilon>0$. If there exists $P_m\succ0$ such that, with
$P_{mn}=P_m\otimes P_n$,
\begin{equation}\label{eq:NL:practical_LMI}
    (1+\varepsilon)X^\top P_{mn}X
    +
    (1+\varepsilon^{-1})Y^\top P_{mn}Y
    -
    P_{mn}
    \prec0,
\end{equation}
\end{subequations}
then the structured LMI \eqref{eq:NL:structured_Lyap} holds, and hence
$\mathcal{S}$ is Schur stable.
\end{corollary}

We next incorporate the norm-bounded uncertainty in
\eqref{eq:AdpId:source} and derive a robust structured LMI that preserves
the Schur stability of $\mathcal S$.

\begin{corollary}[Robust structured LMI]\label{cor:NL:robust_S}
Let $A_0=A_\ast+\Delta A$ where $\|\Delta A\|_2\le a$. 
Define
\begin{equation*}
    \mathcal S_\ast
    \coloneqq
    \Omega\mathbf S
    +
    (\mathbb A_m\otimes I_n)\mathbf A_\ast
    \qquad \textrm{and} \qquad
    \Delta\mathcal S
    \coloneqq
    (\mathbb A_m\otimes I_n)\Delta\mathbf A,
\end{equation*}
where $\mathbf A_\ast=I_m\otimes A_\ast$ and $\Delta\mathbf A=I_m\otimes\Delta A.$
Thus, $\mathcal S=\mathcal S_\ast+\Delta\mathcal S.$

Fix $P_n\succ0$ and $\varepsilon>0$. If there exist
$P_m\succ0$ and $\bar p>0$ such that, with
$P_{mn}\coloneqq P_m\otimes P_n$,
\begin{subequations}\label{eq:NL:robust_LMI}
\begin{align}
    P_{mn}
    &\preceq
    \bar p I_{mn},
    \label{eq:NL:robust_LMI:a}\\
    (1+\varepsilon)\mathcal S_\ast^\top
    P_{mn}\mathcal S_\ast
    +
    (1+\varepsilon^{-1})
    \bar p\,a^2\|\mathbb A_m\|_2^2 I_{mn}
    -P_{mn}
    &\prec0,
    \label{eq:NL:robust_LMI:b}
\end{align}
\end{subequations}
then $\mathcal S$ is Schur stable for every admissible
$\Delta A$ satisfying $\|\Delta A\|_2\le a$.
\end{corollary}

The results of this subsection establish verifiable conditions under which the coupling operator $\mathcal{S}$ is Schur stable despite the network interconnections and model uncertainty. With this stability property in place, we next develop the distributed adaptive laws and analyze the closed-loop stability of the resulting adaptive estimation scheme.

\subsection{Closed-Loop Stability Under Adaptive Laws}\label{subsec:AdpId:Stability}

With the Schur stability of the operator $\mathcal{S}$ established, we next analyze the complete adaptive network. We begin by specifying the adaptive laws, then establish a joint Lyapunov result for the closed-loop error and adaptive parameters. Two corollaries follow: the boundedness of \eqref{eq:AdpId:par_error} and ISS robustness.

Recall the predictor mismatch, captured by
\begin{subequations}
\begin{equation}\label{eq:AdpId:eta}
    \bar\eta_k \coloneqq \Xi_{k}\bar{\xi}_{k} =  \Phi_k \bar z_k + \Psi_k \mathbf g(\Bar{z}_k) + \mathbf{B}_0\bar r_k,
\end{equation}
in \eqref{eq:AdpId:errordyn} and define the network-weighted error direction
\begin{equation}\label{eq:AdpId:zeta}
    \zeta_k \coloneqq \Omega^\top P \bar e_k \in \mathbb{R}^{mn}, \quad \textrm{where } P\succ 0, \Omega = (\mathbb{L}\!\otimes\! I_n).
\end{equation}
\end{subequations}
Each agent implements a local normalization
\begin{equation}\label{eq:AdpId:normalizer}
    N_{i,k} = \max\{\varphi_i, \|z_{i,k}\|^2 + \|g(z_{i,k})\|^2 + \|B_0\|^2\},
\end{equation}
for some $\varphi_i>0$ and the global normalization operator is given by 
$\mathbf{N}_k \coloneqq \diag\{N_{1,k},\dots,N_{m,k}\}$.
The adaptive laws are implemented locally at each agent as
\begin{subequations}\label{eq:AdpId:updates}
\begin{align}
    \Phi_{i,k+1}
    &=
    \Phi_{i,k}
    -
    \frac{\gamma_{A_i}}{N_{i,k}}
    \zeta_{i,k}z_{i,k}^\top,
    \label{eq:AdpId:update_Phi}\\
    \Psi_{i,k+1}
    &=
    \Psi_{i,k}
    -
    \frac{\gamma_{C_i}}{N_{i,k}}
    \zeta_{i,k}g(z_{i,k})^\top,
    \label{eq:AdpId:update_Psi}\\
    r_{i,k+1}
    &=
    r_{i,k}
    -
    \frac{\gamma_{u_i}}{N_{i,k}}
    B_0^\top\zeta_{i,k},
    \label{eq:AdpId:update_r}
\end{align}
\end{subequations}
for $i=1,\dots,m$, where
$\gamma_{A_i},\gamma_{C_i},\gamma_{u_i}>0$
are scalar adaptation gains. The corresponding network quantities and
global adaptation gain matrices are
\begin{equation*}
\begin{aligned}
    \Phi_k   &= \diag\{\Phi_{1,k},\dots,\Phi_{m,k}\},
    &\qquad
    \Gamma_A &= \diag\{\gamma_{A_1},\dots,\gamma_{A_m}\},\\
    \Psi_k   &= \diag\{\Psi_{1,k},\dots,\Psi_{m,k}\},
    &
    \Gamma_C &= \diag\{\gamma_{C_1},\dots,\gamma_{C_m}\},\\
    \Bar r_k &= [r_{1,k}^\top,\dots,r_{m,k}^\top]^\top,
    &
    \Gamma_u &= \diag\{\gamma_{u_1},\dots,\gamma_{u_m}\}.
\end{aligned}
\end{equation*}

We are now ready to establish the main stability statement of the closed-loop error dynamics in \eqref{eq:AdpId:errordyn}.

\begin{theorem}\label{thm:AdpId:CL}
Suppose that $\mathcal{S}$ is Schur stable. Then, for any
$Q=Q^\top\succ0$, there exists $P=P^\top\succ0$ satisfying
$\mathcal{S}^\top P\mathcal{S}-P=-Q.$
Consider the error dynamics \eqref{eq:AdpId:errordyn} under the adaptive
laws \eqref{eq:AdpId:updates}. Then the adaptation gains
$\Gamma_A,\Gamma_C,\Gamma_u\succ0$ and the constants $\varphi_i>0$ in
\eqref{eq:AdpId:normalizer} can be selected such that:

\begin{enumerate}[leftmargin=*]
    \item[i.] The Lyapunov difference satisfies
    \begin{equation}\label{eq:AdpId:Vdiff_main}
        V_{k+1}-V_k
        \le
        -\alpha\|\Bar e_k\|^2
        +c_\eta\|\Bar\eta_k\|^2
        +\beta\|\Bar\delta_{0,k}\|^2,
    \end{equation}
    for some constants $\alpha,c_\eta,\beta>0$, uniformly in $k$.

    \item[ii.] In the disturbance-free case
    $\Bar\delta_{0,k}\equiv\mathbf0_{mn}$, if
    $\sum_{k=0}^{\infty}\|\Bar\eta_k\|^2<\infty,$
    then $\lim_{k\to\infty}\|\Bar e_k\|=\lim_{k\to\infty}\|\Bar\epsilon_k\|=0.$
    Moreover, the adaptive parameter increments satisfy
    \begin{equation}\label{eq:AdpId:param_conv}
        \lim_{k\to\infty}\|\Delta\Phi_k\|_F
        =
        \lim_{k\to\infty}\|\Delta\Psi_k\|_F
        =
        \lim_{k\to\infty}\|\Delta\Bar r_k\|
        =0.
    \end{equation}
\end{enumerate}
\end{theorem}

\begin{corollary}\label{cor:AdpId:ParamBound}
Under the conditions of Theorem~\ref{thm:AdpId:CL}(ii), the adaptive
parameter sequences $(\Phi_k,\Psi_k,\Bar r_k)$ remain bounded for all $k$.
\end{corollary}

\begin{remark}[Persistent excitation]\label{rem:PE}
    Exact identification of the unknown parameters $A_0$ and $C_0$ requires the locally available nonlinear regressor $g(z_{i,k})$ to be sufficiently rich, satisfying the standard persistent excitation \emph{(PE)} condition. Without PE, Theorem~\ref{thm:AdpId:CL} and Corollary~\ref{cor:AdpId:ParamBound} guarantee boundedness of the adaptive parameters, but not necessarily their convergence to the true values.
\end{remark}

\begin{remark}[Cooperative excitation \cite{R18}]\label{rem:CE}
    Since the local regressor is constructed from the aggregated estimate
    $z_{i,k}=\sum\nolimits_{j\in\mathcal{N}_{i}} w_{ij}x_{j,k}$,
    where $\mathcal{N}_{i}\subseteq\mathcal{V}$,
    the proposed observer naturally exploits spatial information from neighboring nodes.
    Consequently, the adopted PE condition is closely related to the cooperative excitation framework in distributed adaptive estimation, where sufficient excitation is provided collectively by the network rather than by individual sensing nodes alone.
\end{remark}

\begin{corollary}[ISS robustness]\label{cor:AdpId:ISS}
Under the assumptions of Theorem~\ref{thm:AdpId:CL}, suppose that
$\|\Bar\eta_k\|\le\eta^\ast$ and $\|\Bar\delta_{0,k}\|\le\delta^\ast$
for all $k$. Then there exist constants
$\vartheta_{e,\eta},\vartheta_{e,\delta},
\vartheta_{\epsilon,\eta},\vartheta_{\epsilon,\delta}>0$,
independent of $k$, such that
\begin{subequations}\label{eq:AdpId:ISS}
\begin{align}
    \limsup_{k\to\infty}\|\Bar e_k\|
    &\le
    \vartheta_{e,\eta}\eta^\ast
    +
    \vartheta_{e,\delta}\delta^\ast,\\
    \limsup_{k\to\infty}\|\Bar\epsilon_k\|
    &\le
    \vartheta_{\epsilon,\eta}\eta^\ast
    +
    \vartheta_{\epsilon,\delta}\delta^\ast.
\end{align}
\end{subequations}
\end{corollary}

Theorem~\ref{thm:AdpId:CL} shows that the composite energy $V_k$ decays in the error component while keeping the adaptive parameters bounded. Corollary~\ref{cor:AdpId:ParamBound} establishes boundedness of the adaptive parameters, while Corollary~\ref{cor:AdpId:ISS} establishes ISS robustness of both the estimation and source-tracking errors under bounded disturbances.

\section{Numerical Results}
\label{sec:NR}

\begin{figure}[h!]
    \centering
    \scalebox{0.90}{{\begin{tikzpicture}
            \centering
            \Text[x=0,y=1.5]{$\mathcal{G}_{\mathrm{s}}\coloneqq$ Star};
            \Vertex[x=0,y=0,label=$0$,color=red,opacity=0.1,size=.5]{L}
            \Vertex[x=1,y=1,label=$1$,color=green,opacity=0.1,size=.5]{1}
            \Vertex[x=-1,y=1,label=$2$,color=green,opacity=0.1,size=.5]{2}
            \Vertex[x=-1,y=-1,label=$3$,color=green,opacity=0.1,size=.5]{3}
            \Vertex[x=1,y=-1,label=$4$,color=green,opacity=0.1,size=.5]{4}
            \Edge[Direct,color=red,label=$1.0$](L)(1)
            \Edge[Direct,color=red,label=$1.0$](L)(2)
            \Edge[Direct,color=red,label=$1.0$](L)(3)
            \Edge[Direct,color=red,label=$1.0$](L)(4)
        \end{tikzpicture}}}
    \qquad
    \scalebox{0.90}{{\begin{tikzpicture}
            \centering
            \Text[x=0,y=1.5]{$\mathcal{G}_{\mathrm{c}}\coloneqq$ Cyclic};
            \Vertex[x=0,y=0,label=$0$,color=red,opacity=0.1,size=.5]{L}
            \Vertex[x=1,y=1,label=$1$,color=green,opacity=0.1,size=.5]{1}
            \Vertex[x=-1,y=1,label=$2$,color=green,opacity=0.1,size=.5]{2}
            \Vertex[x=-1,y=-1,label=$3$,color=green,opacity=0.1,size=.5]{3}
            \Vertex[x=1,y=-1,label=$4$,color=green,opacity=0.1,size=.5]{4}
            \Edge[Direct,color=red,label=$0.4$](L)(1)
            \Edge[Direct,color=red,label=$0.4$](L)(2)
            \Edge[Direct,color=red,label=$0.4$](L)(3)
            \Edge[Direct,color=red,label=$0.4$](L)(4)
            \draw[<->,ultra thick, latex' -latex'] (1) -- node[midway, fill=white, inner sep=2pt]{\small $0.3$} (2);
            \draw[<->,ultra thick, latex' -latex'] (2) -- node[midway, fill=white, inner sep=2pt]{\small $0.3$} (3);
            \draw[<->,ultra thick, latex' -latex'] (3) -- node[midway, fill=white, inner sep=2pt]{\small $0.3$} (4);
            \draw[<->,ultra thick, latex' -latex'] (4) -- node[midway, fill=white, inner sep=2pt]{\small $0.3$} (1);
        \end{tikzpicture}}}
    \qquad
    \scalebox{0.90}{{\begin{tikzpicture}
            \centering
            \Text[x=0,y=1.5]{$\mathcal{G}_{\mathrm{p}}\coloneqq$ Path};
            \Vertex[x=0,y=0,label=$0$,color=red,opacity=0.1,size=.5]{L}
            \Vertex[x=1,y=1,label=$1$,color=green,opacity=0.1,size=.5]{1}
            \Vertex[x=-1,y=1,label=$2$,color=green,opacity=0.1,size=.5]{2}
            \Vertex[x=-1,y=-1,label=$3$,color=green,opacity=0.1,size=.5]{3}
            \Vertex[x=1,y=-1,label=$4$,color=green,opacity=0.1,size=.5]{4}
            \Edge[Direct,color=red,label=$1.0$](L)(1)
            \draw[<->,ultra thick, -latex'] (1) -- node[midway, fill=white, inner sep=2pt]{\small $1.0$} (2);
            \draw[<->,ultra thick, -latex'] (2) -- node[midway, fill=white, inner sep=2pt]{\small $1.0$} (3);
            \draw[<->,ultra thick, -latex'] (3) -- node[midway, fill=white, inner sep=2pt]{\small $1.0$} (4);
        \end{tikzpicture}}}
    \caption{Three network topologies $(\mathcal{G}_{\mathrm{s}}, \mathcal{G}_{\mathrm{c}}, \mathcal{G}_{\mathrm{p}})$ with weights used in the simulations.}
    \label{Fig:topology}
\end{figure}
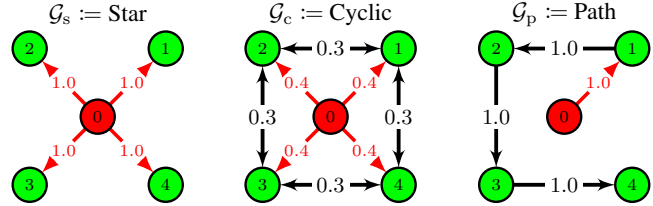

Three directed topologies (star--$\mathcal{G}_{\mathrm{s}}$, cyclic--$\mathcal{G}_{\mathrm{c}}$, path--$\mathcal{G}_{\mathrm{p}}$) are considered, as shown in Fig.~\ref{Fig:topology} along with the weights, under \eqref{eq:ComNet:balance} and Remark~\ref{rem:reachability}. We first test using $m = 4$ sensing nodes, then increase the number of $m$ to verify the time complexity. 

The uncertain nonlinear source evolves as in \eqref{eq:Problem:source} where
\begin{equation*}
    A_0 = \begin{bmatrix} 0.90 & 0.15 \\ -0.05 & 0.95 \end{bmatrix}, \,
    C_0 = \begin{bmatrix} 0.05 & 0 \\ 0.30 & 0.25 \end{bmatrix}, \,
    B_0 = \begin{bmatrix} 0 \\ 0.15 \end{bmatrix}.
\end{equation*}
The nonlinear mapping is $g(x_{0,k}) = [\sin(x^{(1)}_{0,k}), \tanh(x^{(2)}_{0,k})]^\top$, with input $u_{0,k} = 0.7\sin(0.07k) + 0.3\cos(0.05k)$ and disturbance given by
$\delta_{0,k} = 0.05[0.7\sin(0.05k), 0.5\cos(0.09k)]^\top$.

Each node implements \eqref{eq:Problem:sensor}, considering in-neighbor information \eqref{eq:Problem:z}, $N_{i,k}$ (ensuring bounded adaptation even under transient excitation) in \eqref{eq:AdpId:normalizer}, and adaptive laws in \eqref{eq:AdpId:updates}. The initial conditions are $x_{0,0} = [1, 0.5]^\top$, $x_{i,0} \sim \mathbb{N}(0,I_2)$, $\hat A_{i,0} = 0_{2\times2}$, $\hat C_{i,0} = 0_{2\times2}$, and $u_{i,0} = 0$. The normalization floor is $\varphi_i = 10^{-2}$. All simulations are performed for $K=300$ steps.

The proposed distributed adaptive estimator is first evaluated using
$m=4$ sensing nodes over the three communication topologies shown in
Fig.~\ref{Fig:topology}. For the star and cyclic networks, all sensing
nodes employ identical local observer dynamics
$S_i=\mathrm{diag}(0.1,0.1)$ together with uniform adaptation gains
$\Gamma_A=\Gamma_C=\Gamma_u=0.25$. For the path topology, only the first
node has direct access to the source. To compensate for the longer
information propagation path, progressively more contractive local
observer models,
\[
    S_1=0.55I_2,\quad
    S_2=0.45I_2,\quad
    S_3=0.35I_2,\quad
    S_4=0.25I_2,
\]
together with node-dependent adaptation gains
\[
    \Gamma_A=\Gamma_C=\Gamma_u=
    [0.20,\;0.25,\;0.35,\;0.50],
\]
are employed. These parameters satisfy the Schur stability conditions
developed in Section~\ref{sec:AdaptiveId} while preserving stable adaptive
estimation.

\begin{figure*}[t!]
    \centering
    \subfloat[\label{F3a} Star--$\mathcal{G}_{\mathrm{s}}$]{\includegraphics[width=.315\linewidth]{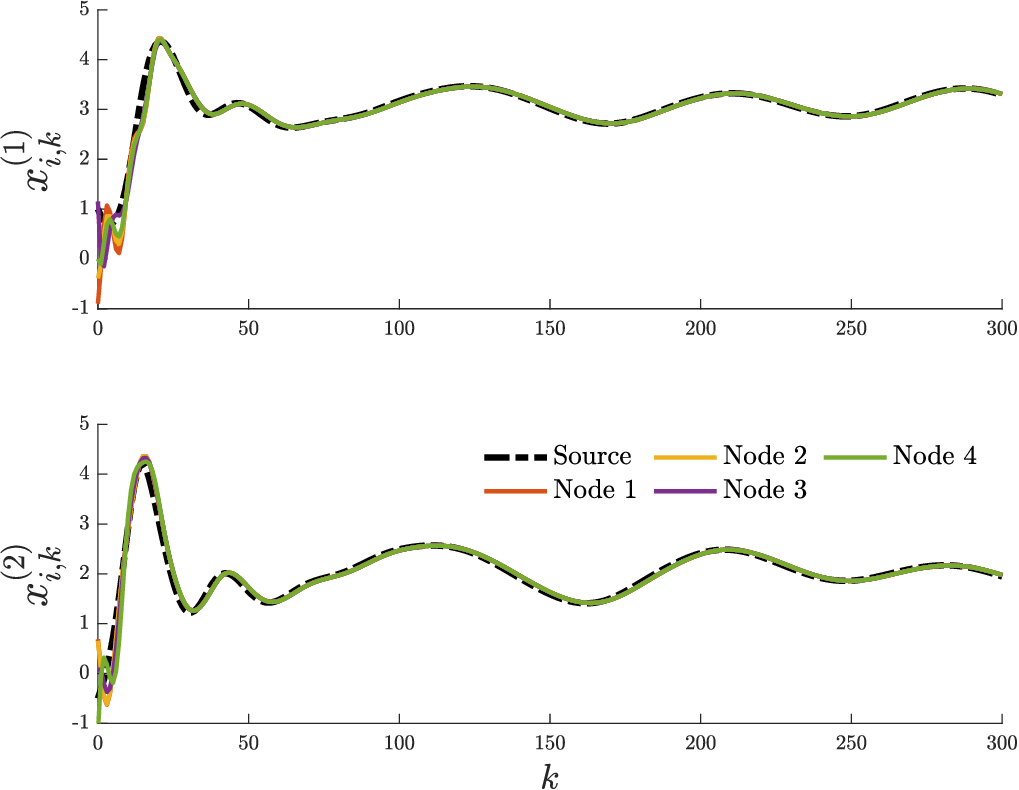}}\quad
    \subfloat[\label{F3b} Cyclic--$\mathcal{G}_{\mathrm{c}}$]{\includegraphics[width=.315\linewidth]{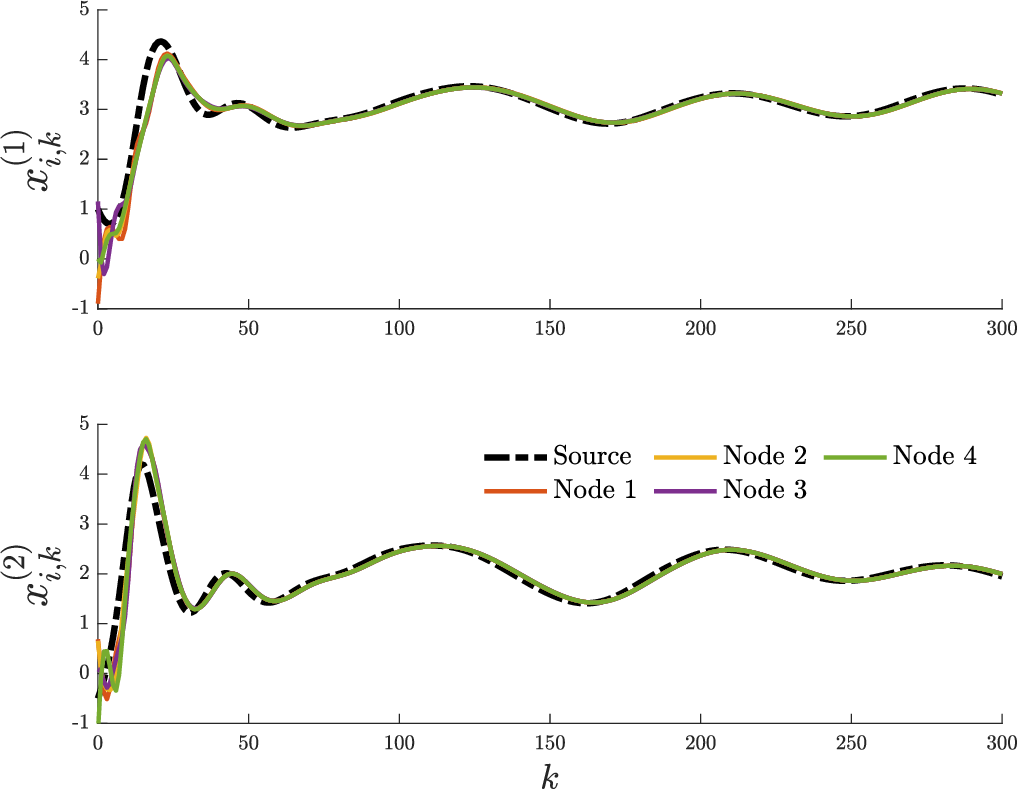}}\quad
    \subfloat[\label{F3c} Path--$\mathcal{G}_{\mathrm{p}}$]{\includegraphics[width=.315\linewidth]{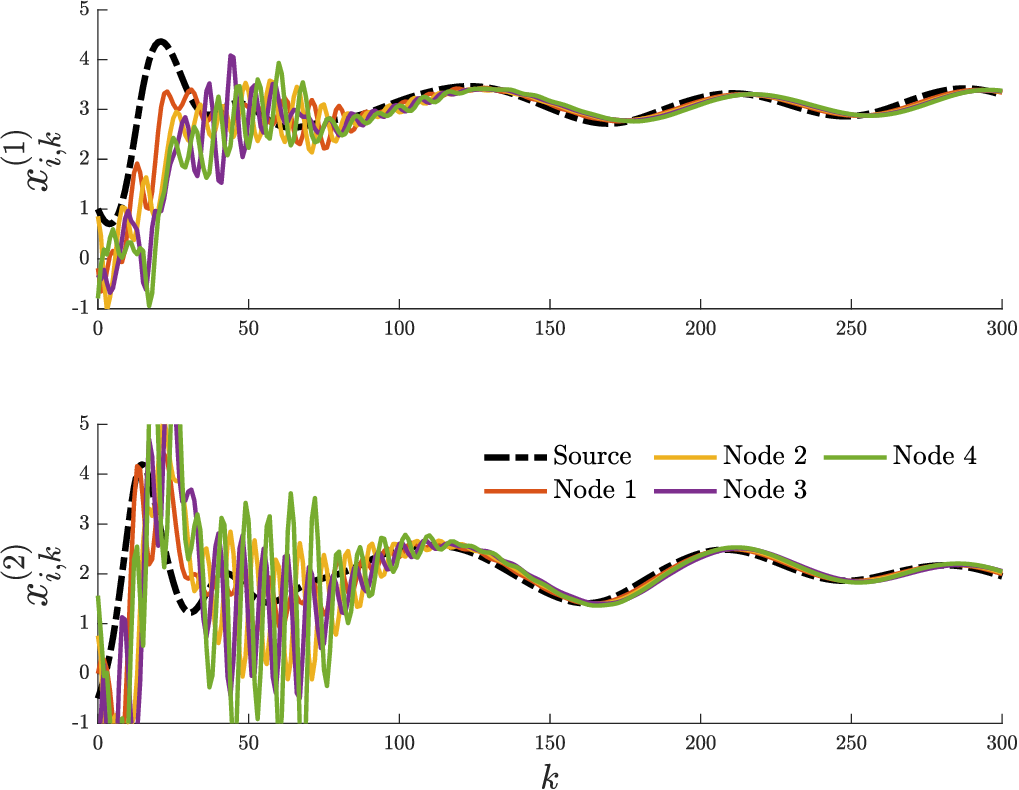}}
    \caption{State trajectories of the distributed adaptive observers for the three communication topologies: (a) star, (b) cyclic, and (c) path. The black dash-dotted curve denotes the true source state, while the colored curves correspond to the four sensing nodes. Despite the different communication structures, all nodes asymptotically track the source state, illustrating the effectiveness of the proposed adaptive estimation scheme.}
    \label{F3}
\end{figure*}

Figure~\ref{F3} illustrates the evolution of the source state together
with the estimated states at the sensing nodes. In all three
communication topologies, every sensing node converges toward the true
source trajectory despite the unknown linear dynamics, nonlinear term,
and external disturbance. This confirms that the proposed distributed
adaptive estimator successfully compensates for the modeling uncertainty
while relying only on local communication.

As expected, the convergence rate depends on the network connectivity.
For the star topology shown in Fig.~\ref{F3a}, every sensing node
receives direct information from the source, resulting in the fastest
convergence and nearly identical trajectories among all nodes. The cyclic
topology in Fig.~\ref{F3b} exhibits a slightly slower transient because
information propagates through neighboring nodes before reaching the
entire network. The path topology in Fig.~\ref{F3c} produces the slowest
transient since only the first sensing node directly observes the
source, requiring information to propagate sequentially through the
network. Nevertheless, all sensing nodes asymptotically track the source
trajectory, demonstrating the robustness of the proposed estimator even
under limited communication.

\begin{figure*}[t!]
    \centering
    \subfloat[\label{F4a} Star--$\mathcal{G}_{\mathrm{s}}$]{\includegraphics[width=.315\linewidth]{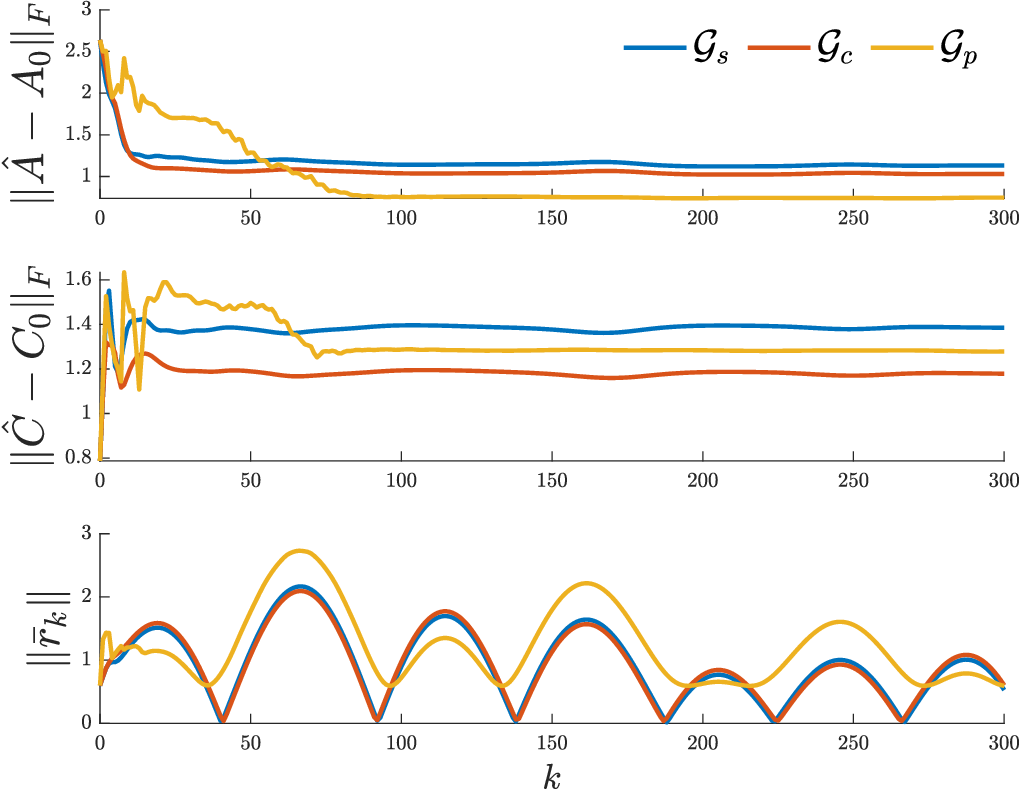}}\quad
    \subfloat[\label{F4b} Cyclic--$\mathcal{G}_{\mathrm{c}}$]{\includegraphics[width=.315\linewidth]{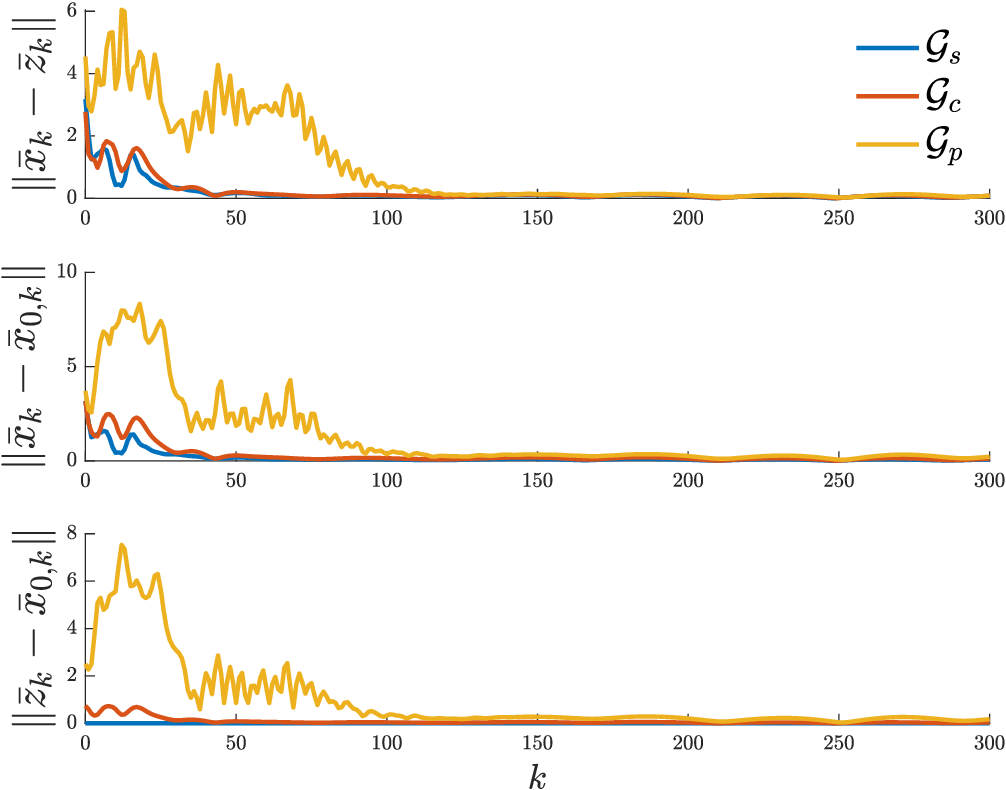}}\quad
    \subfloat[\label{F4c} Path--$\mathcal{G}_{\mathrm{p}}$]{\includegraphics[width=.315\linewidth]{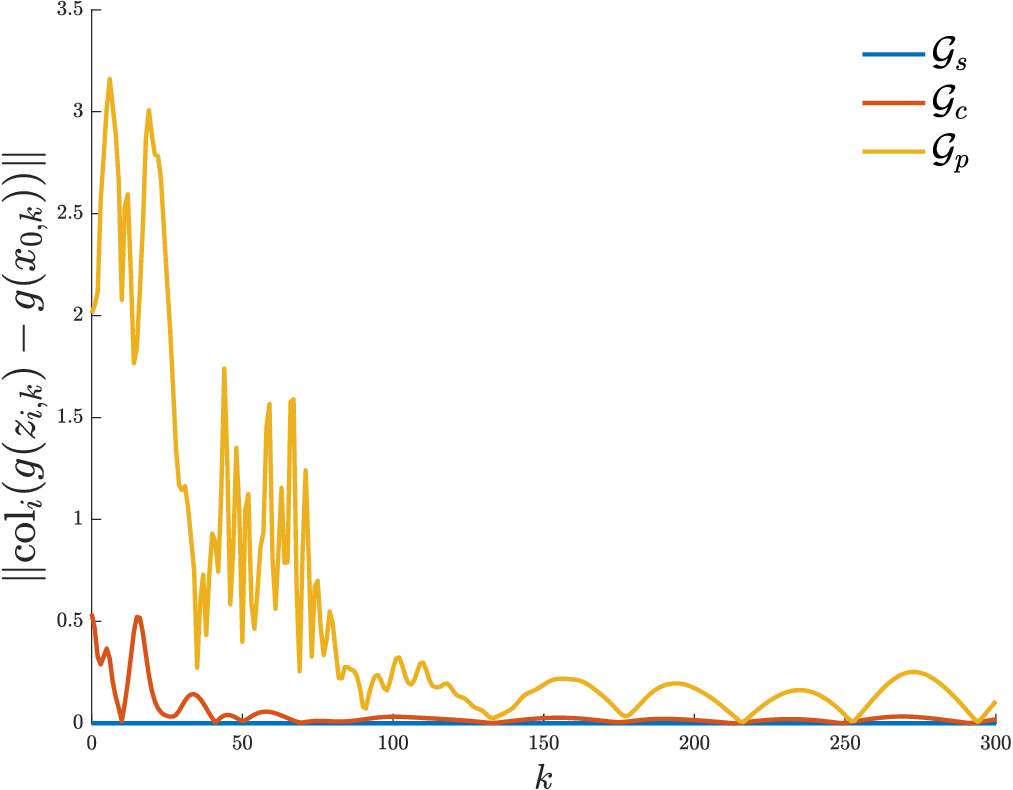}}
    \caption{Evolution of the adaptive parameter estimation errors for the three communication topologies: (a) $\|\hat{A}-A_0\|_F$, (b) $\|\hat{C}-C_0\|_F$, and (c) $\|\bar r_k\|$. The parameter estimates remain bounded and exhibit stable adaptation throughout the simulations, in agreement with the boundedness result of Corollary~\ref{cor:AdpId:ParamBound}.}
    \label{F4}
\end{figure*}

\begin{table}[!t]
    \centering
    \caption{Numerical comparison of the Schur stability conditions for the coupling operator $\mathcal{S}$.}
    \label{tab:S_numerical}
    \begin{tabular}{lcccccc}
    \toprule
    Topology &
    $\rho(\mathcal S)$ &
    Prop.~1 LHS &
    Prop.~1 Margin &
    Lemma~1 Margin &
    Cor.~2 Margin &
    Best $\varepsilon$ \\
    \midrule
    Star
    & 0.1000
    & 0.1000
    & 0.9000
    & 0.2475
    & 0.2475
    & $1.35\times10^{-21}$ \\
    
    Cyclic
    & 0.3484
    & 0.4259
    & 0.5741
    & 0.2357
    & 0.2268
    & $1.87\times10^{-2}$ \\
    
    Path
    & 0.5500
    & 2.0200
    & $-1.0200$
    & 0.1023
    & 0.0952
    & $1.46\times10^{-1}$ \\
    \bottomrule
    \end{tabular}
\end{table}

\begin{table}[!t]
    \centering
    \caption{Comparison of Schur stability certificates for different communication topologies.}
    \label{tab:S_summary}
    \begin{tabular}{lcccc}
    \toprule
    Topology &
    $\rho(\mathcal S)<1$ &
    Proposition~1 &
    Lemma~1 &
    Corollary~2 \\
    \midrule
    Star
    & \checkmark
    & \checkmark
    & \checkmark
    & \checkmark \\
    
    Cyclic
    & \checkmark
    & \checkmark
    & \checkmark
    & \checkmark \\
    
    Path
    & \checkmark
    & $\times$
    & \checkmark
    & \checkmark \\
    \bottomrule
    \end{tabular}
\end{table}

Figure~\ref{F4} reports the evolution of the adaptive parameter
estimation errors. The estimates of $A_0$, $C_0$, and the unknown input
remain bounded throughout the simulations for all communication
topologies. The largest transients occur during the initial adaptation
phase, after which the parameter errors gradually settle to bounded
values. This behavior is consistent with
Corollary~\ref{cor:AdpId:ParamBound}, which guarantees bounded adaptive
parameters. Moreover, the star topology exhibits the fastest adaptation,
followed by the cyclic and path topologies, reflecting the increasing
difficulty of information propagation across the communication network.

Tables~\ref{tab:S_numerical} and~\ref{tab:S_summary} compare the Schur
stability certificates proposed in Section~\ref{sec:AdaptiveId}. Since
Corollary~\ref{cor:NL:practical_LMI} is obtained directly from
Lemma~\ref{lem:NL:LMI_S} through a Young-type relaxation, we report only
the original structured LMI (Lemma~\ref{lem:NL:LMI_S}) and its robust
extension (Corollary~\ref{cor:NL:robust_S}), which are the practically
relevant conditions. The actual coupling operator satisfies
$\rho(\mathcal{S})<1$ for all three communication topologies, confirming
that the corresponding distributed observers are Schur stable.
Proposition~\ref{prop:NL:S_schur} successfully certifies the star and
cyclic networks but fails for the path topology because the norm-based
condition~\eqref{eq:NL:S_schur:cond} is violated. In particular, the
left-hand side of~\eqref{eq:NL:S_schur:cond} equals $2.02>1$, although
the actual spectral radius remains
$\rho(\mathcal{S})=0.55<1$. This illustrates the conservatism of the
simple induced-norm bound.

In contrast, both the structured Lyapunov certificate of
Lemma~\ref{lem:NL:LMI_S} and the simplified structured condition of
Corollary~\ref{cor:NL:practical_LMI} successfully certify Schur
stability for all three communication topologies. These results
demonstrate that preserving the matrix structure of the coupling
operator leads to substantially less conservative stability conditions
than those obtained from induced norm inequalities alone.

\begin{table}[h!]
    \centering
    \caption{Runtime $t_{\mathrm{tot}}$(s) vs.\ number of sensing nodes $m$.}
    \begin{tabular}{c|ccc|c|ccc}
        \toprule
        $m$ & Star & Cyclic & Path & $m$ & Star & Cyclic & Path \\
        \midrule
        50 & 0.145 & 0.146 & 0.139 & 250 & 0.798 & 0.804 & 0.818 \\
        100 & 0.297 & 0.306 & 0.301 & 500 & 1.902 & 2.115 & 2.230 \\
        \bottomrule
    \end{tabular}
    \label{tab:timing}
\end{table}

To evaluate scalability, the proposed estimator is further tested on
networks with
$m\in\{50,100,250,500\}$ sensing nodes across all three communication
topologies. Table~\ref{tab:timing} reports the total runtime
$t_{\mathrm{tot}}$ for $K=300$ sampling steps. The runtime increases
approximately linearly with the number of sensing nodes, which agrees
with the $O(m)$ computational complexity of the proposed distributed
adaptive estimation algorithm. Furthermore, the runtimes are nearly
identical across the three communication topologies for a fixed network
size, indicating that the computational cost is dominated by local
adaptive updates and neighborhood aggregation rather than by the global
network structure. These results demonstrate the scalability of the
proposed estimator to large-scale distributed sensing networks.

\section{Conclusion and Future Work}

This paper presented a discrete-time distributed adaptive estimation framework for nonlinear source systems over directed networks. Each sensing node locally estimated the source state and unknown dynamics using neighbor information and normalized adaptive updates. A Lyapunov-based analysis established input-to-state stability, guaranteeing bounded adaptive parameters and asymptotic convergence of estimation errors under suitable conditions. The proposed framework decouples network coupling and nonlinear dynamics through a Kronecker-structured Lyapunov formulation, yielding explicit spectral and LMI-based conditions that relate graph connectivity, local model stability, and adaptation gains. The normalization mechanism enhances robustness against transient excitation and sampling effects in the discrete-time setting. Simulations on star, cyclic, and path topologies demonstrated accurate source tracking despite partial model knowledge, while computational benchmarks confirmed near-linear scalability with respect to the network size, making the approach suitable for moderately large sensing networks.

Future work will investigate extensions to switching and time-varying communication topologies, communication delays and packet losses, and distributed adaptive estimation under quantized or event-triggered communications. Another promising direction is the integration of adaptive estimation with distributed control and resilient estimation in the presence of malicious or faulty sensing nodes \cite{Wafi-ACC24,R19,R20,Wafi-ACC26,R21}.

\bibliographystyle{IEEEtran}    
\bibliography{EN-Bib.bib}

\section*{Appendix: Proofs}
    \paragraph*{Proof of Proposition~\ref{prop:NL:S_schur}:}
    Using the triangle inequality, submultiplicativity, and the
    Kronecker-product norm identity $\|A\otimes B\|_2=\|A\|_2\|B\|_2,$
    we obtain
    \begin{align}
        \|\mathcal{S}\|_2
        &\le
        \|(\mathbb{L}\otimes I_n)\mathbf{S}\|_2
        +
        \|(\mathbb{A}_m\otimes I_n)\mathbf{A}_0\|_2
        \nonumber\\
        &\le
        \|\mathbb{L}\otimes I_n\|_2\|\mathbf{S}\|_2
        +
        \|\mathbb{A}_m\otimes I_n\|_2\|\mathbf{A}_0\|_2
        \nonumber\\
        &=
        \|\mathbb{L}\|_2\|\mathbf{S}\|_2
        +
        \|\mathbb{A}_m\|_2\|\mathbf{A}_0\|_2.
    \end{align}
    Since $\mathbf{S}=\diag\{S_1,\dots,S_m\}$ and $\mathbf{A}_0=I_m\otimes A_0$,
    we have $\|\mathbf{S}\|_2=\max_i\|S_i\|_2$ and $\|\mathbf{A}_0\|_2=\|A_0\|_2.$
    Therefore, under condition~\eqref{eq:NL:S_schur:cond},
    \[
        \|\mathcal{S}\|_2
        \le
        \|\mathbb{L}\|_2\max_i\|S_i\|_2
        +
        \|\mathbb{A}_m\|_2\|A_0\|_2
        <1.
    \]
    Finally, because $\rho(\mathcal{S})\le\|\mathcal{S}\|_2,$ 
    it follows that $\rho(\mathcal{S})<1,$
    and hence $\mathcal{S}$ is Schur stable.

    \paragraph*{Proof of Lemma~\ref{lem:NL:LMI_S}:}
    Since $P_m\succ0$ and $P_n\succ0$, their Kronecker product satisfies
    $P_{mn}=P_m\otimes P_n\succ0$. Applying the Schur complement to
    \eqref{eq:NL:structured_LMI} gives 
    \begin{equation*}
        P_{mn}-\bigl(P_{mn}\mathcal{S}\bigr)^\top P_{mn}^{-1}\bigl(P_{mn}\mathcal{S}\bigr)\succ0.
    \end{equation*}
    Since $P_{mn}^{-1}P_{mn}=I_{mn}$, it follows that $P_{mn}-\mathcal{S}^\top P_{mn}\mathcal{S}\succ0,$
    which is equivalent to $\mathcal{S}^\top P_{mn}\mathcal{S}-P_{mn}\prec0$ in \eqref{eq:NL:structured_Lyap}.
    
    Now consider $V_k=\Bar e_k^\top P_{mn}\Bar e_k.$
    Along the nominal error dynamics $\Bar e_{k+1}=\mathcal{S}\Bar e_k,$
    we obtain
    \begin{equation*}
        V_{k+1}-V_k
        =
        \Bar e_k^\top
        \bigl(
            \mathcal{S}^\top P_{mn}\mathcal{S}
            -
            P_{mn}
        \bigr)
        \Bar e_k.
    \end{equation*}
    By \eqref{eq:NL:structured_Lyap}, $V_{k+1}-V_k<0, \forall\Bar e_k\neq0.$
    Therefore, the origin of
    $\Bar e_{k+1}=\mathcal{S}\Bar e_k$ is exponentially stable, which is
    equivalent to $\rho(\mathcal{S})<1.$ Hence, $\mathcal{S}$ is Schur stable.

    \paragraph*{Proof of Corollary~\ref{cor:NL:practical_LMI}:}
    Since $\mathcal{S}=X+Y$,
    \begin{subequations}
    \begin{equation}
        \begin{aligned}
        \mathcal{S}^\top P_{mn}\mathcal{S}-P_{mn}
        &= (X+Y)^\top P_{mn}(X+Y)-P_{mn} \\
        &= X^\top P_{mn}X
        +X^\top P_{mn}Y
        +Y^\top P_{mn}X
        +Y^\top P_{mn}Y
        -P_{mn}.
        \end{aligned}
        \label{eq:NL:S_expand}
    \end{equation}
    From \eqref{eq:NL:S_expand}, the mixed terms
    $X^\top P_{mn}Y+Y^\top P_{mn}X$ prevent direct verification of the Lyapunov inequality.
    To obtain a tractable sufficient condition, these terms are upper bounded
    using the weighted matrix Young inequality.
    For any $\varepsilon>0$, the weighted matrix Young inequality gives
    \begin{equation}\label{eq:NL:young}
        X^\top P_{mn}Y
        +
        Y^\top P_{mn}X
        \preceq
        \varepsilon X^\top P_{mn}X
        +\varepsilon^{-1}Y^\top P_{mn}Y.
    \end{equation}
    The inequality is obtained by exploiting the nonnegativity of a suitable
    quadratic form
    \footnote{The choice of the quadratic form is not arbitrary. It is obtained by extending the scalar proof of Young's inequality to matrices. Indeed, for scalars $x,y\in\mathbb{R}$ and any $\varepsilon>0$, the identity $(\sqrt{\varepsilon}\,x-\varepsilon^{-1/2}y)^2\ge0$ expands to $\varepsilon x^2-2xy+\varepsilon^{-1}y^2\ge0$, which is equivalent to $2xy\le\varepsilon x^2+\varepsilon^{-1}y^2$. The matrix inequality is obtained by replacing the scalar variables with $x=P_{mn}^{1/2}X$ and $y=P_{mn}^{1/2}Y$, yielding the quadratic form used above. Thus, the construction is simply the matrix analogue of completing the square.}.
    Specifically, since $P_{mn}\succ0$, its unique symmetric
    square root $\sqrt{P_{mn}}$ exists, and for any $\varepsilon>0$,
    \[
        \bigl(
            \sqrt{\varepsilon}\,P_{mn}^{\frac{1}{2}}X
            -
            \varepsilon^{-\frac{1}{2}}P_{mn}^{\frac{1}{2}}Y
        \bigr)^\top
        \bigl(
            \sqrt{\varepsilon}\,P_{mn}^{\frac{1}{2}}X
            -
            \varepsilon^{-\frac{1}{2}}P_{mn}^{\frac{1}{2}}Y
        \bigr)
        \succeq0.
    \]
    Expanding the above expression and rearranging terms yields
    \eqref{eq:NL:young}.
    Next, substituting \eqref{eq:NL:young} into \eqref{eq:NL:S_expand} yields
    \begin{equation}
        \begin{aligned}
        \mathcal{S}^\top P_{mn}\mathcal{S}-P_{mn}
        &=
        X^\top P_{mn}X
        +X^\top P_{mn}Y
        +Y^\top P_{mn}X
        +Y^\top P_{mn}Y
        -P_{mn} \\
        &\preceq
        X^\top P_{mn}X
        +\varepsilon X^\top P_{mn}X
        +\varepsilon^{-1}Y^\top P_{mn}Y
        +Y^\top P_{mn}Y
        -P_{mn} \\
        &=
        (1+\varepsilon)X^\top P_{mn}X
        +(1+\varepsilon^{-1})Y^\top P_{mn}Y
        -P_{mn}.
        \end{aligned}
    \end{equation}
    \end{subequations}
    Therefore, if \eqref{eq:NL:practical_LMI} holds, then
    $\mathcal{S}^\top P_{mn}\mathcal{S}-P_{mn}\prec0.$
    Lemma~\ref{lem:NL:LMI_S} then implies that $\mathcal{S}$ is Schur stable.
    
    \paragraph*{Proof of Corollary~\ref{cor:NL:robust_S}:}
    Since $\mathcal S=\mathcal S_\ast+\Delta\mathcal S,$ we have
    \begin{subequations}
    \begin{equation}\label{eq:NL:robust_expand}
        \begin{aligned}
        \mathcal{S}^\top P_{mn}\mathcal{S}-P_{mn}
        &= (\mathcal S_\ast+\Delta\mathcal S)^\top P_{mn}(\mathcal S_\ast+\Delta\mathcal S)-P_{mn} \\
        &=\mathcal S_\ast^\top P_{mn}\mathcal S_\ast
        +\mathcal S_\ast^\top P_{mn}\Delta\mathcal S
        +\Delta\mathcal S^\top P_{mn}\mathcal S_\ast
        +\Delta\mathcal S^\top P_{mn}\Delta\mathcal S
        -P_{mn}.
        \end{aligned}
    \end{equation}
    Applying the weighted Young inequality gives
    \begin{align}
        \mathcal S_\ast^\top P_{mn}\Delta\mathcal S
        +
        \Delta\mathcal S^\top P_{mn}\mathcal S_\ast
        &\preceq
        \varepsilon
        \mathcal S_\ast^\top P_{mn}\mathcal S_\ast
        +
        \varepsilon^{-1}
        \Delta\mathcal S^\top P_{mn}\Delta\mathcal S.
        \label{eq:NL:robust_young}
    \end{align}
    Substituting \eqref{eq:NL:robust_young} into
    \eqref{eq:NL:robust_expand} yields
    \begin{equation}\label{eq:NL:robust_bound1}
        \begin{aligned}
        \mathcal{S}^\top P_{mn}\mathcal{S}-P_{mn}
        &\preceq
        \mathcal S_\ast^\top P_{mn}\mathcal S_\ast
        +\varepsilon\mathcal S_\ast^\top P_{mn}\mathcal S_\ast
        +\varepsilon^{-1}\Delta\mathcal S^\top P_{mn}\Delta\mathcal S
        +\Delta\mathcal S^\top P_{mn}\Delta\mathcal S
        -P_{mn} \\
        &\preceq
        (1+\varepsilon)
        \mathcal S_\ast^\top P_{mn}\mathcal S_\ast
        +\left(1+\varepsilon^{-1}\right)\Delta\mathcal S^\top P_{mn}\Delta\mathcal S
        -P_{mn}.
        \end{aligned}
    \end{equation}
    From \eqref{eq:NL:robust_LMI:a}, $P_{mn}\preceq\bar p I_{mn},$
    and hence
    \begin{align}
        \Delta\mathcal S^\top P_{mn}\Delta\mathcal S
        &\preceq
        \bar p\,
        \Delta\mathcal S^\top\Delta\mathcal S
        \nonumber\\
        &\preceq
        \bar p\,
        \|\Delta\mathcal S\|_2^2 I_{mn}.
        \label{eq:NL:robust_bound2}
    \end{align}
    Moreover, by the Kronecker-product norm identity,
    \begin{align}
        \|\Delta\mathcal S\|_2
        &=
        \|(\mathbb A_m\otimes I_n)
        (I_m\otimes\Delta A)\|_2
        \nonumber\\
        &=
        \|\mathbb A_m\otimes\Delta A\|_2
        \nonumber\\
        &=
        \|\mathbb A_m\|_2\|\Delta A\|_2
        \nonumber\\
        &\le
        a\|\mathbb A_m\|_2.
        \label{eq:NL:robust_deltaS}
    \end{align}
    Therefore,
    \[
        \Delta\mathcal S^\top P_{mn}\Delta\mathcal S
        \preceq
        \bar p\,a^2\|\mathbb A_m\|_2^2 I_{mn}.
    \]
    Using this bound in \eqref{eq:NL:robust_bound1} gives
    \begin{align}
        \mathcal S^\top P_{mn}\mathcal S-P_{mn}
        &\preceq
        (1+\varepsilon)
        \mathcal S_\ast^\top P_{mn}\mathcal S_\ast
        +
        (1+\varepsilon^{-1})
        \bar p\,a^2\|\mathbb A_m\|_2^2 I_{mn}
        -
        P_{mn}.
    \end{align}
    \end{subequations}
    Condition \eqref{eq:NL:robust_LMI:b} therefore implies
    $\mathcal S^\top P_{mn}\mathcal S-P_{mn}\prec0$
    for every admissible $\Delta A$. By
    Lemma~\ref{lem:NL:LMI_S}, $\mathcal S$ is Schur stable.

    \paragraph*{Proof of Theorem~\ref{thm:AdpId:CL}:}
    Recall the estimation-error dynamics
    \begin{subequations}
    \begin{equation}\label{eq:pf:error_dynamics}
        \Bar e_{k+1}
        =
        \mathcal S\Bar e_k
        +
        \Omega\Bar\eta_k
        +
        \Omega\mathbf C_0\Delta\mathbf g_k
        -
        \Omega\Bar\delta_{0,k},
    \end{equation}
    where $\Bar\eta_k$ is defined in \eqref{eq:AdpId:eta}. Let
    $\mathbf N_k\coloneqq\diag\{N_{1,k},\dots,N_{m,k}\}$
    with $\underline\varphi\coloneqq\min_i\varphi_i>0,$
    so that
    \[
        \mathbf N_k^{-1}\preceq\underline\varphi^{-1}I_m.
    \]
    Also recall $\zeta_k=\Omega^\top P\Bar e_k,$
    and let $P=P^\top\succ0$ satisfy $\mathcal S^\top P\mathcal S-P=-Q,$ where
    $Q=Q^\top\succ0.$
    
    Consider the Lyapunov function
    \begin{equation}\label{eq:pf:Vk}
        V_k
        =
        \Bar e_k^\top P\Bar e_k
        +
        \tr\!\left(
            \Phi_k^\top\Gamma_A^{-1}\Phi_k
        \right)
        +
        \tr\!\left(
            \Psi_k^\top\Gamma_C^{-1}\Psi_k
        \right)
        +
        \Bar r_k^\top\Gamma_u^{-1}\Bar r_k.
    \end{equation}
    For compactness, define
    \begin{equation}\label{eq:pf:wk}
        \Bar w_k
        \coloneqq
        \mathbf C_0\Delta\mathbf g_k
        -
        \Bar\delta_{0,k}.
    \end{equation}
    Then \eqref{eq:pf:error_dynamics} becomes
    \[
        \Bar e_{k+1}
        =
        \mathcal S\Bar e_k
        +
        \Omega\Bar\eta_k
        +
        \Omega\Bar w_k.
    \]
    
    Using the adaptive laws \eqref{eq:AdpId:updates}, the Lyapunov
    difference satisfies
    \begin{align}
        \Delta V_k
        &\coloneqq
        V_{k+1}-V_k
        \nonumber\\
        &=
        -\Bar e_k^\top Q\Bar e_k
        +
        2\Bar e_k^\top\mathcal S^\top P\Omega\Bar\eta_k
        +
        \Bar\eta_k^\top\Omega^\top P\Omega\Bar\eta_k
        \nonumber\\
        &\quad
        -
        2\zeta_k^\top
        (\mathbf N_k^{-1}\otimes I_n)\Bar\eta_k
        +
        \mathcal R_k
        \nonumber\\
        &\quad
        +
        2\Bar e_k^\top\mathcal S^\top P\Omega\Bar w_k
        +
        2\Bar\eta_k^\top\Omega^\top P\Omega\Bar w_k
        +
        \Bar w_k^\top\Omega^\top P\Omega\Bar w_k,
        \label{eq:pf:Vdiff_raw}
    \end{align}
    where the quadratic terms produced by the discrete adaptive updates are
    \begin{align}
        \mathcal R_k
        \coloneqq
        \sum_{i=1}^{m}
        \Bigg[
        &\frac{\gamma_{A_i}}{N_{i,k}^{2}}
        \|\zeta_{i,k}z_{i,k}^{\top}\|_{F}^{2}
        +
        \frac{\gamma_{C_i}}{N_{i,k}^{2}}
        \|\zeta_{i,k}g(z_{i,k})^{\top}\|_{F}^{2}
        +
        \frac{\gamma_{u_i}}{N_{i,k}^{2}}
        \|B_0^{\top}\zeta_{i,k}\|^{2}
        \Bigg].
        \label{eq:pf:Rk}
    \end{align}
    
    Let $\Bar\gamma\coloneqq\max_i\{\gamma_{A_i},\gamma_{C_i},\gamma_{u_i}\}.$
    From the definition of the local normalizer in
    \eqref{eq:AdpId:normalizer}, one obtains
    \[
        N_{i,k}
        \ge
        \|z_{i,k}\|^2+\|g(z_{i,k})\|^2+\|B_0\|^2,
    \]
    and therefore
    \[
        \frac{\|z_{i,k}\|^2}{N_{i,k}}\le1,\qquad
        \frac{\|g(z_{i,k})\|^2}{N_{i,k}}\le1,\qquad
        \frac{\|B_0\|^2}{N_{i,k}}\le1.
    \]
    Since $N_{i,k}\ge\underline\varphi$, it follows that
    \begin{equation}\label{eq:pf:Rk_bound}
        \mathcal R_k
        \le
        \frac{\Bar\gamma}{\underline\varphi}
        \|\zeta_k\|^2.
    \end{equation}
    Next, because $\zeta_k^\top\Bar\eta_k=\Bar e_k^\top P\Omega\Bar\eta_k,$
    we have
    \begin{align}
        \Bar e_k^\top\mathcal S^\top P\Omega\Bar\eta_k
        &=
        \zeta_k^\top\Bar\eta_k
        +
        \Bar e_k^\top
        (\mathcal S^\top P-P)\Omega\Bar\eta_k.
        \label{eq:pf:cross_decomp}
    \end{align}
    Substituting \eqref{eq:pf:cross_decomp} into
    \eqref{eq:pf:Vdiff_raw} gives
    \begin{align}
        \Delta V_k
        &=
        -\Bar e_k^\top Q\Bar e_k
        +
        2\zeta_k^\top
        \left(
            I_{mn}
            -
            \mathbf N_k^{-1}\otimes I_n
        \right)
        \Bar\eta_k
        \nonumber\\
        &\quad
        +
        2\Bar e_k^\top
        (\mathcal S^\top P-P)\Omega\Bar\eta_k
        +
        \Bar\eta_k^\top\Omega^\top P\Omega\Bar\eta_k
        \nonumber\\
        &\quad
        +
        \mathcal R_k
        +
        2\Bar e_k^\top\mathcal S^\top P\Omega\Bar w_k
        \nonumber\\
        &\quad
        +
        2\Bar\eta_k^\top\Omega^\top P\Omega\Bar w_k
        +
        \Bar w_k^\top\Omega^\top P\Omega\Bar w_k.
        \label{eq:pf:Vdiff_mid}
    \end{align}
    
    Since $\mathbf N_k$ is diagonal, define
    \begin{equation*}
        \begin{aligned}
        \mathbf{D}_k
        &\coloneqq
        I_{mn}
        -
        \mathbf N_k^{-1}\otimes I_n \\
        &=
        \diag\left\{
        \left(1-\frac{1}{N_{1,k}}\right)I_n,\dots,
        \left(1-\frac{1}{N_{m,k}}\right)I_n
        \right\},
        \end{aligned}
    \end{equation*}
    and therefore
    \[
        \|\mathbf{D}_k\|_2
        =\max_i\left|1-\frac{1}{N_{i,k}}\right|.
    \]
    Since $N_{i,k}\ge\underline\varphi$, the quantity
    $|1-N_{i,k}^{-1}|$ is bounded by its value at
    $N_{i,k}=\underline\varphi$ or its limiting value $1$ as
    $N_{i,k}\to\infty$. Hence,
    \begin{equation*}
        \|\mathbf{D}_k\|_2
        \le
        c_N
        \coloneqq
        \max\left\{
            1,\,
            \left|1-\underline\varphi^{-1}\right|
        \right\}.
    \end{equation*}
    Using the Cauchy--Schwarz inequality and the bound
    $\|\mathbf D_k\|_2\le c_N$, we obtain
    \begin{align*}
        2\zeta_k^\top\mathbf D_k\Bar\eta_k
        &\le
        2\bigl|\zeta_k^\top\mathbf D_k\Bar\eta_k\bigr|
        \nonumber\\
        &\le
        2\|\zeta_k\|\,\|\mathbf D_k\Bar\eta_k\|
        \nonumber\\
        &\le
        2\|\mathbf D_k\|_2\|\zeta_k\|\,\|\Bar\eta_k\|
        \nonumber\\
        &\le
        2c_N\|\zeta_k\|\,\|\Bar\eta_k\|.
    \end{align*}
    Applying Young's inequality
    $2ab\le\varepsilon a^2+\varepsilon^{-1}b^2$
    with $a=\|\zeta_k\|$, $b=c_N\|\Bar\eta_k\|$, and
    $\varepsilon=\varepsilon_N>0$ yields
    \begin{align}
        2\zeta_k^\top\mathbf D_k\Bar\eta_k
        &\le
        \varepsilon_N\|\zeta_k\|^2
        +
        \frac{c_N^2}{\varepsilon_N}
        \|\Bar\eta_k\|^2.
        \label{eq:pf:normalization_bound}
    \end{align}
    Similarly, for any $\varepsilon_e>0$,
    \begin{align}
        2\Bar e_k^\top
        (\mathcal S^\top P-P)\Omega\Bar\eta_k
        \le
        \varepsilon_e\|\Bar e_k\|^2
        +
        \frac{
            \|(\mathcal S^\top P-P)\Omega\|_2^2
        }{\varepsilon_e}
        \|\Bar\eta_k\|^2,
        \label{eq:pf:e_eta_bound}
    \end{align}
    and
    \begin{equation}\label{eq:pf:eta_quad_bound}
        \Bar\eta_k^\top\Omega^\top P\Omega\Bar\eta_k
        \le
        \|\Omega^\top P\Omega\|_2
        \|\Bar\eta_k\|^2.
    \end{equation}
    
    The terms involving $\Bar w_k$ satisfy, for any
    $\varepsilon_w,\varepsilon_\eta>0$,
    \begin{align}
        2\Bar e_k^\top\mathcal S^\top P\Omega\Bar w_k
        &\le
        \varepsilon_w\|\Bar e_k\|^2
        +
        \frac{\|\mathcal S^\top P\Omega\|_2^2}
        {\varepsilon_w}
        \|\Bar w_k\|^2,
        \label{eq:pf:e_w_bound}\\
        2\Bar\eta_k^\top\Omega^\top P\Omega\Bar w_k
        &\le
        \varepsilon_\eta\|\Bar\eta_k\|^2
        +
        \frac{\|\Omega^\top P\Omega\|_2^2}
        {\varepsilon_\eta}
        \|\Bar w_k\|^2,
        \label{eq:pf:eta_w_bound}\\
        \Bar w_k^\top\Omega^\top P\Omega\Bar w_k
        &\le
        \|\Omega^\top P\Omega\|_2
        \|\Bar w_k\|^2.
        \label{eq:pf:w_quad_bound}
    \end{align}
    Define
    \begin{equation}\label{eq:pf:cw}
        c_w
        \coloneqq
        \frac{\|\mathcal S^\top P\Omega\|_2^2}
        {\varepsilon_w}
        +
        \frac{\|\Omega^\top P\Omega\|_2^2}
        {\varepsilon_\eta}
        +
        \|\Omega^\top P\Omega\|_2.
    \end{equation}
    
    By the Lipschitz bound established before the theorem,
    \begin{equation*}
        \|\Delta\mathbf g_k\|
        \le
        \kappa_g\|\Bar e_k\|,
    \end{equation*}
    where $\kappa_g\coloneqq L_g\|\mathbb A_m\mathbb L^{-1}\otimes I_n\|_2.$
    Consequently,
    \[
        \|\mathbf C_0\Delta\mathbf g_k\|
        \le
        \kappa_C\|\Bar e_k\|,
    \]
    where $\kappa_C\coloneqq\|\mathbf C_0\|_2\kappa_g.$
    Using \eqref{eq:pf:wk}, for any $\varepsilon_\delta>0$,
    \begin{align}
        \|\Bar w_k\|^2
        &=
        \|
            \mathbf C_0\Delta\mathbf g_k
            -
            \Bar\delta_{0,k}
        \|^2
        \nonumber\\
        &\le
        (1+\varepsilon_\delta)\kappa_C^2
        \|\Bar e_k\|^2
        +
        (1+\varepsilon_\delta^{-1})
        \|\Bar\delta_{0,k}\|^2.
        \label{eq:pf:w_bound}
    \end{align}
    Moreover, from $\zeta_k=\Omega^\top P\Bar e_k,$ we have
    \begin{equation}\label{eq:pf:zeta_bound}
        \|\zeta_k\|^2
        \le
        c_\zeta\|\Bar e_k\|^2,
    \end{equation}
    where $c_\zeta\coloneqq\|\Omega^\top P\|_2^2.$
    Substituting
    \eqref{eq:pf:Rk_bound}--\eqref{eq:pf:zeta_bound}
    into \eqref{eq:pf:Vdiff_mid} yields
    \begin{equation}\label{eq:pf:Vdiff_final}
        \Delta V_k
        \le
        -\alpha\|\Bar e_k\|^2
        +
        c_\eta\|\Bar\eta_k\|^2
        +
        \beta\|\Bar\delta_{0,k}\|^2,
    \end{equation}
    where
    \begin{align*}
        \alpha
        &\coloneqq
        \lambda_{\min}(Q)
        -
        \varepsilon_e
        -
        \varepsilon_w
        -
        \left(
            \varepsilon_N
            +
            \frac{\Bar\gamma}{\underline\varphi}
        \right)c_\zeta
        -
        c_w(1+\varepsilon_\delta)\kappa_C^2,\\
        c_\eta
        &\coloneqq
        \frac{c_N^2}{\varepsilon_N}
        +
        \frac{
            \|(\mathcal S^\top P-P)\Omega\|_2^2
        }{\varepsilon_e}
        +
        \|\Omega^\top P\Omega\|_2
        +
        \varepsilon_\eta,\\
        \beta
        &\coloneqq
        c_w(1+\varepsilon_\delta^{-1}).
    \end{align*}
    Suppose that the adaptation gains and positive Young parameters are
    chosen such that
    \begin{equation}\label{eq:pf:small_gain}
        \alpha>0,
    \end{equation}
    where $\alpha$ is defined above. This proves item~(i).
    
    For item~(ii), suppose that $\Bar\delta_{0,k}\equiv\mathbf0_{mn}$
    and $\sum_{k=0}^{\infty}\|\Bar\eta_k\|^2<\infty.$
    Then \eqref{eq:pf:Vdiff_final} gives
    \[
        V_{k+1}
        \le
        V_k
        -
        \alpha\|\Bar e_k\|^2
        +
        c_\eta\|\Bar\eta_k\|^2.
    \]
    Summing from $k=0$ to $K$ yields
    \begin{align}
        \alpha
        \sum_{k=0}^{K}\|\Bar e_k\|^2
        \le
        V_0-V_{K+1}
        +
        c_\eta
        \sum_{k=0}^{K}\|\Bar\eta_k\|^2
        \le
        V_0
        +
        c_\eta
        \sum_{k=0}^{\infty}\|\Bar\eta_k\|^2.
    \end{align}
    Therefore,
    \[
        \sum_{k=0}^{\infty}\|\Bar e_k\|^2<\infty,
    \]
    which implies $\lim_{k\to\infty}\|\Bar e_k\|=0.$
    Since $\Bar\epsilon_k=(\mathbb L^{-1}\otimes I_n)\Bar e_k,$
    we also obtain
    \[
        \lim_{k\to\infty}\|\Bar\epsilon_k\|=0.
    \]
    Finally, the normalized update laws imply
    \begin{align}
        \|\Delta\Phi_{i,k}\|_F
        &\le
        \frac{\gamma_{A_i}}{\sqrt{\underline\varphi}}
        \|\zeta_{i,k}\|,\\
        \|\Delta\Psi_{i,k}\|_F
        &\le
        \frac{\gamma_{C_i}}{\sqrt{\underline\varphi}}
        \|\zeta_{i,k}\|,\\
        \|\Delta r_{i,k}\|
        &\le
        \frac{\gamma_{u_i}}{\sqrt{\underline\varphi}}
        \|\zeta_{i,k}\|.
    \end{align}
    \end{subequations}
    Because $\zeta_k=\Omega^\top P\Bar e_k\to0,$
    it follows that
    \[
        \lim_{k\to\infty}\|\Delta\Phi_k\|_F
        =
        \lim_{k\to\infty}\|\Delta\Psi_k\|_F
        =
        \lim_{k\to\infty}\|\Delta\Bar r_k\|
        =
        0,
    \]
    establishing \eqref{eq:AdpId:param_conv}. This proves item~(ii).

    \paragraph*{Proof of Corollary~\ref{cor:AdpId:ParamBound}:}
    Under the conditions of Theorem~\ref{thm:AdpId:CL}(ii),
    $\Bar\delta_{0,k}\equiv\mathbf0_{mn}$ and
    $\sum_{k=0}^{\infty}\|\Bar\eta_k\|^2<\infty$.
    From \eqref{eq:AdpId:Vdiff_main},
    \[
        V_{k+1}
        \le
        V_k
        -
        \alpha\|\Bar e_k\|^2
        +
        c_\eta\|\Bar\eta_k\|^2.
    \]
    Summing from $k=0$ to $K$ and dropping the nonpositive error term gives
    \[
        V_{K+1}
        \le
        V_0
        +
        c_\eta
        \sum_{k=0}^{K}\|\Bar\eta_k\|^2
        \le
        V_0
        +
        c_\eta
        \sum_{k=0}^{\infty}\|\Bar\eta_k\|^2
        <\infty.
    \]
    Hence, $V_k$ is uniformly bounded. Since the parameter-error terms
    $\tr(\Phi_k^\top\Gamma_A^{-1}\Phi_k)$,
    $\tr(\Psi_k^\top\Gamma_C^{-1}\Psi_k)$, and
    $\Bar r_k^\top\Gamma_u^{-1}\Bar r_k$
    are nonnegative components of $V_k$, the sequences
    $(\Phi_k,\Psi_k,\Bar r_k)$ remain bounded.

    \paragraph*{Proof of Corollary~\ref{cor:AdpId:ISS}:}
    Consider the error Lyapunov function
    \[
        W_k
        \coloneqq
        \Bar e_k^\top P\Bar e_k.
    \]
    Repeating the error-state part of the proof of
    Theorem~\ref{thm:AdpId:CL}, together with the Lipschitz bound on
    $\Delta\mathbf g_k$, yields
    \begin{equation}\label{eq:pf:ISS_W}
        W_{k+1}-W_k
        \le
        -\alpha_W\|\Bar e_k\|^2
        +
        c_{\eta,W}\|\Bar\eta_k\|^2
        +
        c_{\delta,W}\|\Bar\delta_{0,k}\|^2,
    \end{equation}
    for some constants
    $\alpha_W,c_{\eta,W},c_{\delta,W}>0$.
    Since $W_k\le\lambda_{\max}(P)\|\Bar e_k\|^2,$
    we have
    \[
        \|\Bar e_k\|^2
        \ge
        \frac{W_k}{\lambda_{\max}(P)}.
    \]
    Hence,
    \[
        W_{k+1}
        \le
        (1-\mu)W_k
        +
        c_{\eta,W}\|\Bar\eta_k\|^2
        +
        c_{\delta,W}\|\Bar\delta_{0,k}\|^2,
    \]
    where
    \[
        \mu
        \coloneqq
        \frac{\alpha_W}{\lambda_{\max}(P)}.
    \]
    Choosing the design parameters such that
    $0<\mu<1$, and using the assumptions, $\|\Bar\eta_k\|\le\eta^\ast$ and $\|\Bar\delta_{0,k}\|\le\delta^\ast$,
    give
    \[
        W_{k+1}
        \le
        (1-\mu)W_k
        +
        c_{\eta,W}\eta^{\ast2}
        +
        c_{\delta,W}\delta^{\ast2}.
    \]
    Iterating this recursion and letting $k\to\infty$ yields
    \[
        \limsup_{k\to\infty}W_k
        \le
        \frac{
            c_{\eta,W}\eta^{\ast2}
            +
            c_{\delta,W}\delta^{\ast2}
        }{\mu}.
    \]
    Since $W_k\ge\lambda_{\min}(P)\|\Bar e_k\|^2,$
    we obtain
    \[
        \limsup_{k\to\infty}\|\Bar e_k\|
        \le
        \sqrt{
            \frac{
                c_{\eta,W}\eta^{\ast2}
                +
                c_{\delta,W}\delta^{\ast2}
            }{
                \mu\lambda_{\min}(P)
            }
        }.
    \]
    Applying
    $\sqrt{a+b}\le\sqrt a+\sqrt b$
    gives
    \begin{subequations}
    \begin{equation}
        \limsup_{k\to\infty}\|\Bar e_k\|
        \le
        \vartheta_{e,\eta}\eta^\ast
        +
        \vartheta_{e,\delta}\delta^\ast,
    \end{equation}
    where
    \[
        \vartheta_{e,\eta}
        \coloneqq
        \sqrt{
            \frac{c_{\eta,W}}
            {\mu\lambda_{\min}(P)}
        },
        \qquad
        \vartheta_{e,\delta}
        \coloneqq
        \sqrt{
            \frac{c_{\delta,W}}
            {\mu\lambda_{\min}(P)}
        }.
    \]
    
    Finally, since $\Bar\epsilon_k=(\mathbb L^{-1}\otimes I_n)\Bar e_k,$
    it follows that
    \[
        \limsup_{k\to\infty}\|\Bar\epsilon_k\|
        \le
        \|\mathbb L^{-1}\otimes I_n\|_2
        \limsup_{k\to\infty}\|\Bar e_k\|,
    \]
    and therefore
    \begin{equation}
        \limsup_{k\to\infty}\|\Bar\epsilon_k\|
        \le
        \vartheta_{\epsilon,\eta}\eta^\ast
        +
        \vartheta_{\epsilon,\delta}\delta^\ast,
    \end{equation}
    \end{subequations}
    where
    \[
        \vartheta_{\epsilon,\eta}
        \coloneqq
        \|\mathbb L^{-1}\otimes I_n\|_2
        \vartheta_{e,\eta},
        \qquad
        \vartheta_{\epsilon,\delta}
        \coloneqq
        \|\mathbb L^{-1}\otimes I_n\|_2
        \vartheta_{e,\delta}.
    \]
    This establishes \eqref{eq:AdpId:ISS}.

\end{document}